\documentclass[%
reprint,
showkeys,
superscriptaddress,
amsmath,amssymb,
aps,
longbibliography,
floatfix,
]{revtex4-2}

\usepackage{graphicx}
\usepackage{dcolumn}
\usepackage{bm}
\usepackage{hyperref}
\usepackage[mathlines]{lineno}
\usepackage{xcolor}
\usepackage{mathtools}
\hypersetup{colorlinks=true,
            linkbordercolor=teal, 
            citecolor=teal,
            urlcolor=teal,
            filecolor=teal,
            linkcolor=teal}
\usepackage{orcidlink}
\definecolor{orcidlogocol}{HTML}{A6CE39}

\begin{document}


\title{Evidence of directional structural superlubricity and\\L{\'e}vy flights in a van der Waals heterostructure}


\author{M. Le Ster \orcidlink{0000-0002-3874-799X}}
\email{maxime.lester@fis.uni.lodz.pl}

\author{P. Krukowski \orcidlink{0000-0002-1368-6908}}
\author{M. Rogala \orcidlink{0000-0002-7898-5087}}
\author{P. Dabrowski}
\author{I. Lutsyk}
\author{K. Toczek}
\author{K. Podlaski}
\affiliation{Faculty of Physics and Applied Informatics, University of Lodz, Pomorska 149/153, Lodz, 90-236, Poland}

\author{T. O. Mente\c{s} \orcidlink{0000-0003-0413-9272
}}
\author{F. Genuzio \orcidlink{0000-0003-0699-2525}}
\author{A. Locatelli \orcidlink{0000-0002-8072-7343}}
\affiliation{Elettra-Sincrotrone Trieste S.C.p.A., Strada Statale 14 - km 163,5 in AREA Science Park, 34149 Basovizza, Trieste, Italy}

\author{G. Bian}
\affiliation{Department of Physics and Astronomy, University of Missouri, Columbia, MO 65211, United States}

\author{T.-C. Chiang \orcidlink{0000-0001-7118-8299}}
\affiliation{Department of Physics, University of Illinois at Urbana-Champaign, Urbana, IL, 61801-3080, United States}

\author{S. A. Brown \orcidlink{0000-0002-6041-4331}}
\affiliation{The MacDiarmid Institute for Advanced Materials and Nanotechnology, University of Canterbury, Christchurch 8140, New Zealand}

\author{P. J. Kowalczyk \orcidlink{0000-0001-6310-4366}}
\email{pawel.kowalczyk@uni.lodz.pl}
\affiliation{Faculty of Physics and Applied Informatics, University of Lodz, Pomorska 149/153, Lodz, 90-236, Poland}

\date{\today}


\begin{abstract}
Structural superlubricity is a special frictionless contact in which two crystals are in incommensurate arrangement such that relative in-plane translation is associated with vanishing energy barrier crossing. So far, it has been realized in multilayer graphene and other van der Waals two-dimensional crystals with hexagonal or triangular crystalline symmetries, leading to isotropic frictionless contacts. Directional structural superlubricity, to date unrealized in two-dimensional systems, is possible when the reciprocal lattices of the two crystals coincide in one direction only. Here, we evidence directional structural superlubricity a $\alpha$-bismuthene/graphite van der Waals system, manifested by spontaneous hopping of the islands over hundreds of nanometres at room temperature, resolved by low-energy electron microscopy and supported by registry simulations. Statistical analysis of individual and collective $\alpha$-bismuthene islands populations reveal a heavy-tailed distribution of the hopping lengths and sticking times indicative of L{\'e}vy flight dynamics, largely unobserved in condensed-matter systems.
\end{abstract}


\maketitle
\newpage

\section{Introduction}

Friction emerges from energy dissipation at the contact interface between two materials in relative motion and is present in virtually all mechanical systems. Friction, responsible for both direct energy losses (essentially, heat dissipation) and indirect costs (performace reduction, material wear and repair) has been estimated to contribute to approximately 23\% of the world's energy consumption \cite{Holmberg2017, Luo2021}. Therefore, it is desirable to investigate interfacial systems that offer a substantial decrease in friction coefficients. Structural superlubricity \cite{Hod2012, Hod2018, Martin2018, Zhai2019, Vanossi2020, Wang2024} is a regime of motion between crystalline solids in incommensurate contact (with no additional lubricant phase) leading to considerable friction coefficient reduction typically below instrumental resolution \cite{Hod2012, Hod2018}. Over the last decade, structural superlubricity has been described in a number of van der Waals (vdW) two-dimensional (2D) systems such as graphene on graphite \cite{Hod2012, Ouyang2018}, bilayer graphene \cite{Androulidakis2020}, multilayer hexagonal boron nitride (hBN) \cite{Hod2012} and molybdenum disulfide (MoS$_2$) on graphite and hBN \cite{Liao2021}.
\begin{figure*}[t]
\centering
\includegraphics[width=\textwidth]{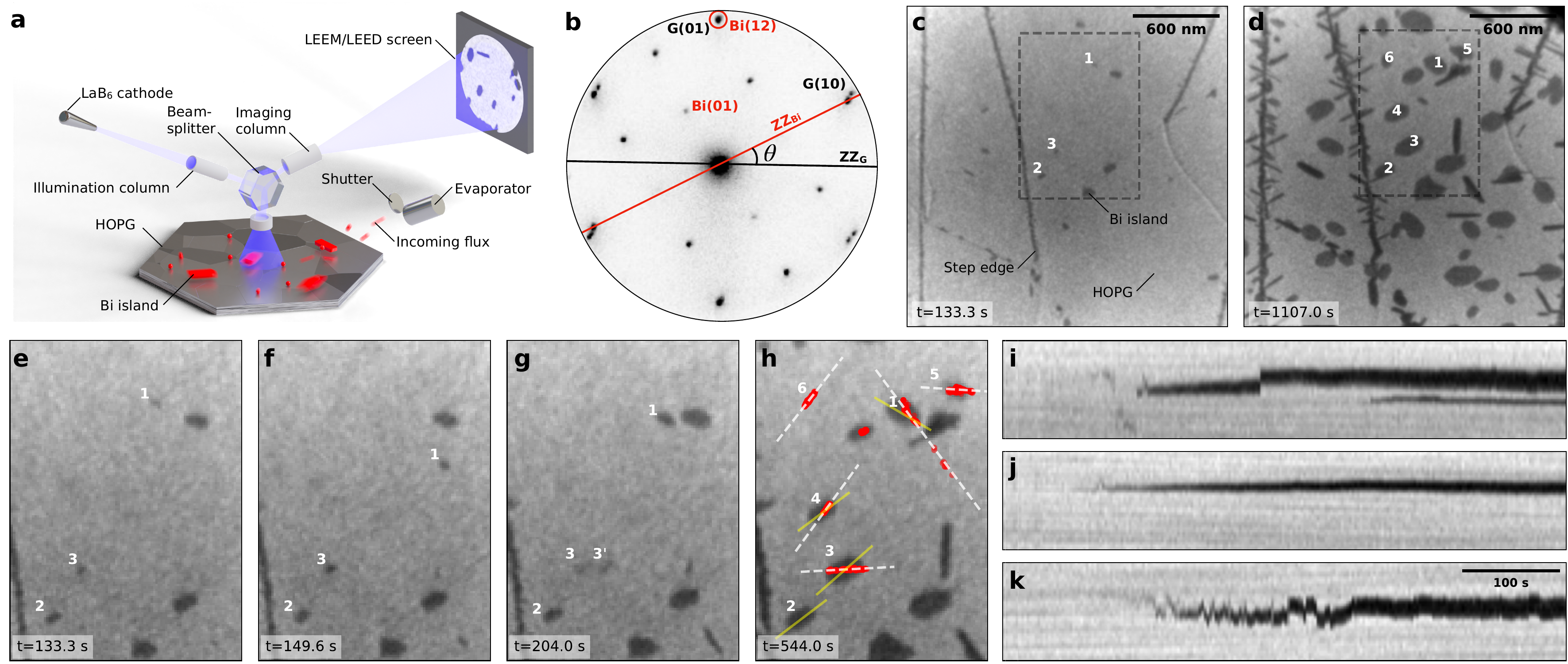}
\caption{\textbf{LEEM and $\mu$-LEED experiments.} (a) Schematics of the experimental setup. (b) $\mu$-LEED pattern ($E=40$~eV) of a single $\alpha$-Bi island. The black and red lines indicate graphite and $\alpha$-Bi zigzag directions, respectively. The twist angle ($\theta\simeq28^\circ$) is indicated. (c) and (d) $2.25\times2.25~\mu$m$^2$ LEEM images recorded during Bi deposition at 133.3 and 1107.0 s, respectively. (e-h) Sequence of LEEM images (area in the dashed squares in (c, d)) showing $\alpha$-Bi growth snapshots recorded at 133.3, 149.6, 204.0 and 544.0~s into deposition. The yellow lines in (h) indicate $\alpha$-Bi elongation (zigzag) direction, and the red dots indicate the position of the centre of mass of the islands during deposition (dashed white lines are linear regressions). (i, j, k) Time-dependent cross-sections of islands 1, 2 and 3 respectively (panel vertical dimension: 500 nm). All measurements in (b-k) were obtained at room temperature.}
\label{fig:1}
\end{figure*}
Despite a considerable progress in the investigation of superlubricity in 2D vdW systems, the reported component crystalline layers almost always possess a triangular or hexagonal surface symmetry (graphene, MoS$_2$, hBN). To the best of our knowledge, the only exception is a theoretical investigation of bilayer $\alpha$-phosphorene \cite{Losi2020}; however no nanoscale systems involving superlubricity with distinct symmetries have been published so far.

In recent years, phenomenological models were developed by Hod \cite{Hod2012, Hod2018} which showed that atomic registry at the crystalline interface is crucial to understanding structural superlubricity. To that effect, registry index (RI) simulations, which track the atomic overlap at the interface, offer qualitative approximations of the translational energy landscape $U(\mathbf{r}_C)$ (with $\mathbf{r}_C$ the relative translation vector). Recently, Panizon \cite{Panizon2023} classified three types of crystalline contacts which depend on the set of coincidence reciprocal lattice vectors $\Omega$ at the crystalline interface ($\Omega$ can be either empty, or form a 1D, or a 2D lattice). The type of contact determines the structure of $U(\mathbf{r}_C)$ which can be corrugated in all directions (type-A), in one direction only (type-B), or uniform (type-C). In type-A contacts, a translation of the adsorbate layer requires overcoming non-zero energy barriers in all translation directions; this regime of motion corresponds to \emph{directional locking} \cite{Trillitzsch2018}, prohibiting structural superlubricity. Type-C contacts on the contrary lead to vanished energy barriers in all translation directions, \emph{i.e.}, structural superlubricity as described in most experimental reports \cite{Hod2012, Hod2018, Martin2018, Ouyang2018, Zhai2019, Androulidakis2020, Vanossi2020, Liao2021}. Lastly, type-B contacts lead to one-dimensional translation energy landscapes such that the adsorbate layer can glide freely along directional tracks (\emph{nanohighways}) associated with quasi-vanishing energy barriers, while other directions lead to much larger friction coefficients, typical of type-A. Type-B is possible only when the two lattices have different rotational symmetries \cite{Panizon2023}, and was demonstrated experimentally in a self-assembled triangular lattice of colloidal particles on a surface with square symmetry (with lattice parameters of the order of several micrometres) \cite{Cao2021, Panizon2023}. However, there is to date no evidence of such directional structural superlubricity regime in a nanoscale system with atomically-clean contact.

In this work, we report the first realization of directional structural superlubricity in a nanoscale 2D system, comprised of self-assembled $\alpha$-bismuthene ($\alpha$-Bi) islands on a highly-ordered pyrolithic graphite (HOPG) substrate. We use low-energy electron microscopy (LEEM) during and after Bi deposition to observe, in real time (at a frame rate up to 0.2 s) and real space, the motion dynamics of $\alpha$-Bi islands, which appear to spontaneously hop back-and-forth (along graphitic zigzag directions) with hopping lengths $\ell$ as large as 600~nm. Our RI simulations support these observations and reveal a unique translational energy potential landscape as a function of the twist angle $\theta$ between $\alpha$-Bi and HOPG, in agreement with the predicted directional superlubricity which enables spontaneous diffusion of the islands at room temperature. Interestingly, the distributions of hopping lengths and sticking times (duration between successive hopping events) $P(\ell)$ and $P(\tau)$ respectively, for both individual islands and for the global population are heavy-tailed (\emph{i.e.}, which decays slower than exponentially; $P(\ell)\sim\ell^{-2.2}$ and $P(\tau)\sim\tau^{-2.3}$), highlighting that $\alpha$-Bi islands on graphite surfaces display L{\'e}vy flights statistics, which have been extremely scarce in solid-state physics, \emph{a fortiori} for large and massive nanostructures comprised of up to hundreds of thousands of atoms.


\section{LEEM and $\mu$-LEED results}


\paragraph{General description of the growth.}$\alpha$-Bi islands were grown on HOPG using a simple atomic deposition process described in depth in previous reports \cite{Scott2005, McCarthy2010, Kowalczyk2011, Kowalczyk2013, LeSter2019, Kowalczyk2020, Salehi2023}. In this work, the growth process is observed using LEEM and micro-spot $\mu$-LEED \cite{Locatelli2006, Mentes2012, Mentes2014}. Figure~\ref{fig:1}(a) illustrates the experimental set-up employed to monitor \emph{in-situ} the Bi deposition on HOPG. Figure~\ref{fig:1}(b) shows the diffraction pattern from a single $\alpha$-Bi island, obtained by restricting the e-beam illumination to a spot size of the order of 500~nm. The diffraction pattern resolves both $\alpha$-Bi and the underlying HOPG substrate giving an estimate of the twist angle $\theta\simeq28^\circ$, in agreement with previous experimental reports \cite{Kowalczyk2015, LeSter2019, Salehi2023}. Interestingly, the $\mu$-LEED pattern shows a unique reciprocal lattice coincidence vector $\mathbf{G}(01)=\mathbf{Bi}(12)$, indicating a directional commensurate matching, which is a crucial factor in terms of anisotropic diffusion \cite{Cao2021, Panizon2023}.
In fact, by symmetry, $\theta=32^\circ$ should lead to a similar situation where the coincidence vector is $\mathbf{G}(\bar{1}1)=\mathbf{Bi}(\bar{1}2)$ (see SI section~\ref{si:coincidencevectors}). Figure~\ref{fig:1}(c,d) shows LEEM images obtained (c) shortly after island nucleation, \emph{i.e.}, at the early stages of growth, with small islands decorating graphite step edges and occasionally located in the middle of the terraces; and (d) after deposition of a Bi dose equivalent to a coverage of $\sim1$~ monolayer (ML). The LEEM image shown in Fig.~\ref{fig:1}(d) is representative of the later stages of $\alpha$-Bi growth on HOPG. Besides step-edge decoration, one can observe two different types of $\alpha$-Bi crystals: \emph{(i)} relatively large and flat islands with thickness in the range 2-4 ML; \emph{(ii)} narrower and elongated nanorods, both elongated along a preferential direction parallel to $\alpha$-Bi $\mathbf{R}_1$ direction analogous to bulk Bi $\langle 1 \bar{1} 0 \rangle$ direction \cite{Scott2005, McCarthy2010, Kowalczyk2011, Kowalczyk2013, LeSter2019, Kowalczyk2020, Salehi2023}. Note that most islands are anchored at steps, which act as nucleation centres in the early growth stages. We occasionally observe island ripening, which is common for two-dimensional systems in which adatom attachment/detachment takes place between the condensed island and the two dimensional lattice gas \cite{Kowalczyk2014}.


\paragraph{Spontaneous hopping.}
Quantitative analysis of LEEM movies shows that, during Bi deposition at room temperature, about $\sim20\%$ of the islands (including ones with areas as large as $20000$~nm$^2$) spontaneously diffuse along straight lines in a back-and-forth fashion, which we refer to as \emph{hopping}. The complete LEEM sequence is shown in supplementary data movie MOV-1 recorded during the entire deposition (about 18 minutes). Several zoomed-in frames are shown in Fig.~\ref{fig:1}(e-h). The hopping behaviour is clearly observed for island 1, captured in four different positions in all panels, in contrast to island 2 which remains stationary during deposition. The centre of mass of islands 1-6 recorded over the whole deposition are overlaid onto Fig.~\ref{fig:1}(h) and confirm the directional character of the hopping, where the $\alpha$-Bi islands diffuse in straight lines (white dashed lines) along directions $\simeq30^\circ$ off their elongation direction (solid yellow lines). Furthermore, the hopping directions of the different islands are rotated by $\pm60^\circ$, suggesting that this phenomenon is governed by graphitic crystallographic directions. The variations within the island population is highlighted by the waterfall plots in Fig.~\ref{fig:1}(i-k). Island 1 diffuses back-and-forth rapidly at the beginning of the deposition and then stabilizes for about 100~s, finally hopping once more and remains stationary for the rest of the deposition. In contrast, island 2 is fixed during the entire deposition, likely due to defect-pinning \cite{Chee2019}. Island 3, on the other hand, appears to shuttle to and from two fixed locations, likely alternating between two shallow pinning states caused by local defects in the HOPG crystalline structure. Closer inspection of islands 3 and 4 in Fig.~\ref{fig:1}(h) shows that despite having nearly parallel elongation axes (and therefore twist angles $\theta$) their spontaneous hopping directions differ significantly ($\simeq 60^\circ$). In fact the two islands appear to have a slightly different twist (yellow lines are not exactly parallel). This unique friction/twist dependence, in which a twist change of several degrees leads to a drastic change the in easy-translation direction, is corroborated below with our simulations.


\paragraph{Origin of hopping.}
In LEEM, the electron energy at the sample surface is very low (few tens of eV) compared to the much higher energies found in transmission electron microscopy (TEM; around 100 keV). Therefore in our case we can rule out electron-induced processes such as cluster diffusion, previously observed in TEM studies \cite{Chee2019, Yesibolati2020}. Instead, we attribute the island diffusion to surface phonons, as was observed for C$_{60}$ molecules on Au(111) surface \cite{Jia2020} and Au clusters on graphite \cite{Vasileiadis2019}. We investigate the influence of the substrate thermal activity by increasing the temperature ($\sim400$~K, see SI section~\ref{si:thermalactivation} and SI MOV-3) which results in the unpinning of several islands previously immobilized at room temperature, further indicating that the momentum transfer has a thermal origin and that pinning/unpinning activation processes occur at defect sites.


\paragraph{Diffusion velocity.}Due to the limited time resolution of the LEEM set-up at high magnifications ($0.2-2.7$ frames per second), estimating the velocity of the $\alpha$-Bi islands during a diffusion event is difficult, due to both under-sampling and the relatively high exposure time used in LEEM (few hundred milliseconds). Island 3 in Fig.~\ref{fig:1}(g) however, captured in two different locations ($3$ and $3'$), allows to make a lower estimate of the diffusion velocity, $v_d = 700$~nm/s. Alternative estimations exploiting the sometimes smeared appearance of the islands (see SI section~\ref{si:velocity}) suggest $v_d = 790 - 1900$~nm/s.


\section{Registry Index Simulations}

\begin{figure*}[t]
\centering
\includegraphics[width=\textwidth]{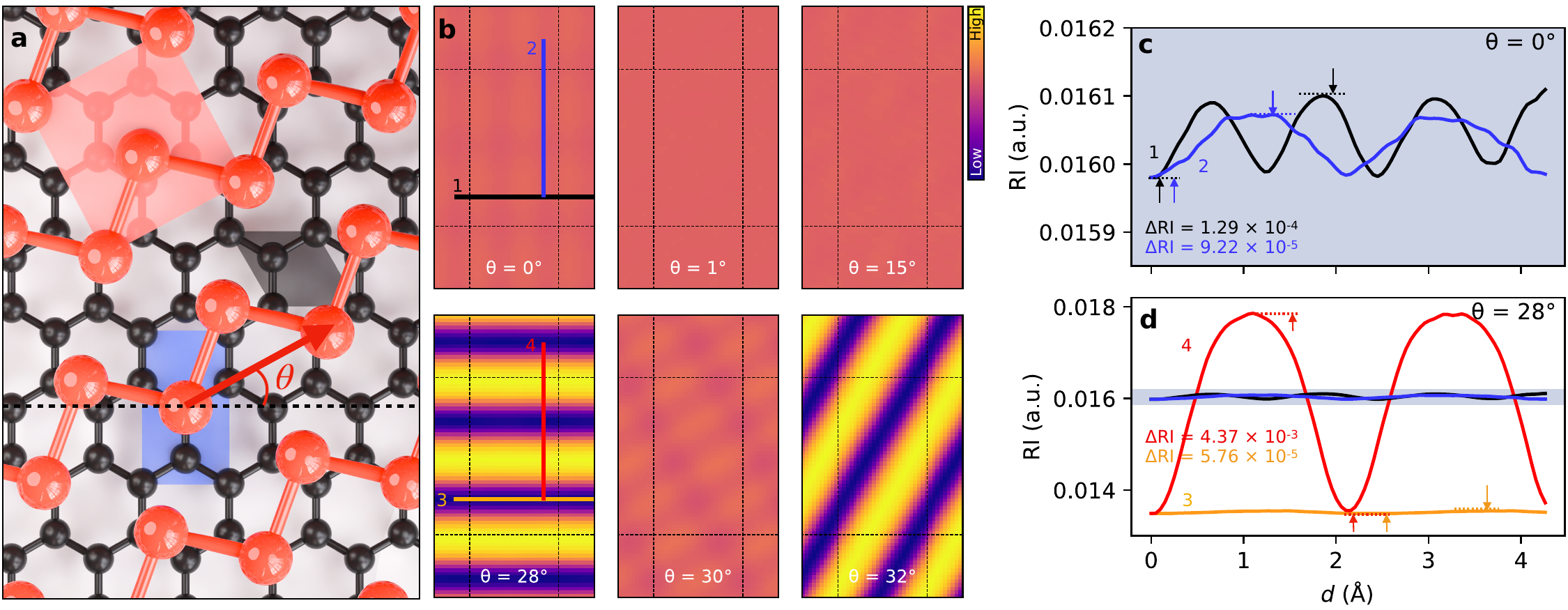}
\caption{\textbf{Registry index simulations.} (a) Ball-and-stick model of the $\alpha$-Bi (red)/graphite (black), with unit-cells shown by semi-transparent regions of the same colors. Only the atomic layers at the interface are shown. The blue area highlights to the translation space of RI maps. The zigzag direction of graphite (dashed line) and $\alpha$-Bi's $\mathbf{R}'_1$ (red arrow) are indicated. (b) RI maps for selected twist angles specified on the RI maps. Dashed lines correspond to borders of the translation space. (c, d) RI profiles taken from solid lines of the same colours in (b), for $\theta=0^\circ$ (c) and $\theta=28^\circ$ (d) (dotted lines are arrow indicate minima and maxima). The profiles in (b-d) are also numbered for clarity.}
\label{fig:2}
\end{figure*}

As evidenced by LEEM, the trajectories of the $\alpha$-Bi islands are rotated by $60^\circ$, suggesting that the hopping tracks follow crystallographic directions. The observed behaviour suggests a partly commensurate superlubricity, where the $\alpha$-Bi adsorbate layer forms a 1D superlattice. This remains in contrast to most commensurate systems defined with two non-colinear commensurate vectors \cite{Panizon2023}. The $\mu$-LEED image in Fig.~\ref{fig:1}(b) shows that the reciprocal lattice vectors $\mathbf{Bi}(12)$ and $\mathbf{G}(01)$ are superposed (and correspond to a unique coincidence vector), as expected for type-B contact leading to directional superlubricity. Figure~\ref{fig:2}(a) shows the crystal structures of $\alpha$-Bi and HOPG at the interface (deeper graphene monolayers, and $\alpha$-Bi above the first atomic layer in contact with graphite are ignored).

To gain insight into the nanoscale friction properties of $\alpha$-Bi/HOPG, we perform RI simulations on finite 2D $\alpha$-Bi flakes for different twist angles. The RI is the overlap area between the substrate and the adsorbate interfacial lattices, using circular domains for each atomic site \cite{Hod2012}. RI maps, obtained by calculating the RI for a range of lateral translation vectors $\mathbf{r}_c$, can approximate $U(\mathbf{r}_c)$ of the adsorbed layer \cite{Hod2012, Hod2018} (more details on the simulations are given in SI section ~\ref{si:ri}). Figure~\ref{fig:2}(b) shows RI maps for an $\alpha$-Bi slab comprising of $30\times30$ unit-cells (1800 atoms) for several twist angles. For several twist values ($\theta = 0, 1, 15$ and $30^\circ$), the potential energy landscape is nearly uniform, meaning that the diffusion barrier is independent of the direction. In contrast, the RI maps obtained for $\theta=28^\circ$ and $\theta=32^\circ$ possess a strong 1D character, consistent with large friction anisotropy. The global RI minima are found for these two twist angles suggesting these are the most energetically favourable configurations (see RI maps obtained for a large range of twist angle in SI Fig.~\ref{fig:si_ri_theta}), in agreement with the observed twist angles in the experiment. The two special twist angles correspond to the two symmetrically equivalent partly commensurate configurations, $\mathbf{Bi}(12) = \mathbf{G}(01)$ and $\mathbf{Bi}(\bar{1}2) = \mathbf{G}(\bar{1}1)$ respectively (see SI section~\ref{si:coincidencevectors} for additional information), corresponding to the energy landscapes of islands~3 and~4 above, consistently with their minor difference in twist. The low RI pathways correspond to the \emph{nanohighways}, demonstrated for microscale colloidal particles \cite{Panizon2023}.

To further characterize the potential energy barriers associated with the translation of $\alpha$-Bi, we consider line profiles across the RI maps to qualitatively compare frictional properties for different twist angles and different hopping directions. Several profiles visible as colour lines in Fig.~\ref{fig:2}(b) are shown in Fig.~\ref{fig:2}(c, d). The RI profiles along graphite's zigzag (black) and armchair (blue) directions are very similar, showing a maximum corrugation of $\Delta\mathrm{RI}=1.29\times10^{-4}$ and $\Delta\mathrm{RI}=9.22\times10^{-5}$ respectively (a ratio of $\simeq1.4$) consistent with isotropic behaviour. In contrast, the equivalent RI profiles along the same translation directions for $\theta=28^\circ$ are characterized with significantly different energy barriers ($\Delta \mathrm{RI}=5.76\times10^{-5}$ and $\Delta\mathrm{RI}=4.73\times10^{-3}$ for profiles along graphite's zigzag and armchair translation, respectively) with a ratio $\simeq76$, consistent with anisotropic friction behaviour. The RI corrugation along the preferential directions is substantially vanished and in fact decreases with the $\alpha$-Bi slab (see SI section~\ref{si:ri} for size-dependent simulations). These unique translational energy potential landscapes are in agreement with theory for type-B contacts \cite{Panizon2023}.

\section{Hopping statistics}

\begin{figure*}[t]
\centering
\includegraphics[width=\textwidth]{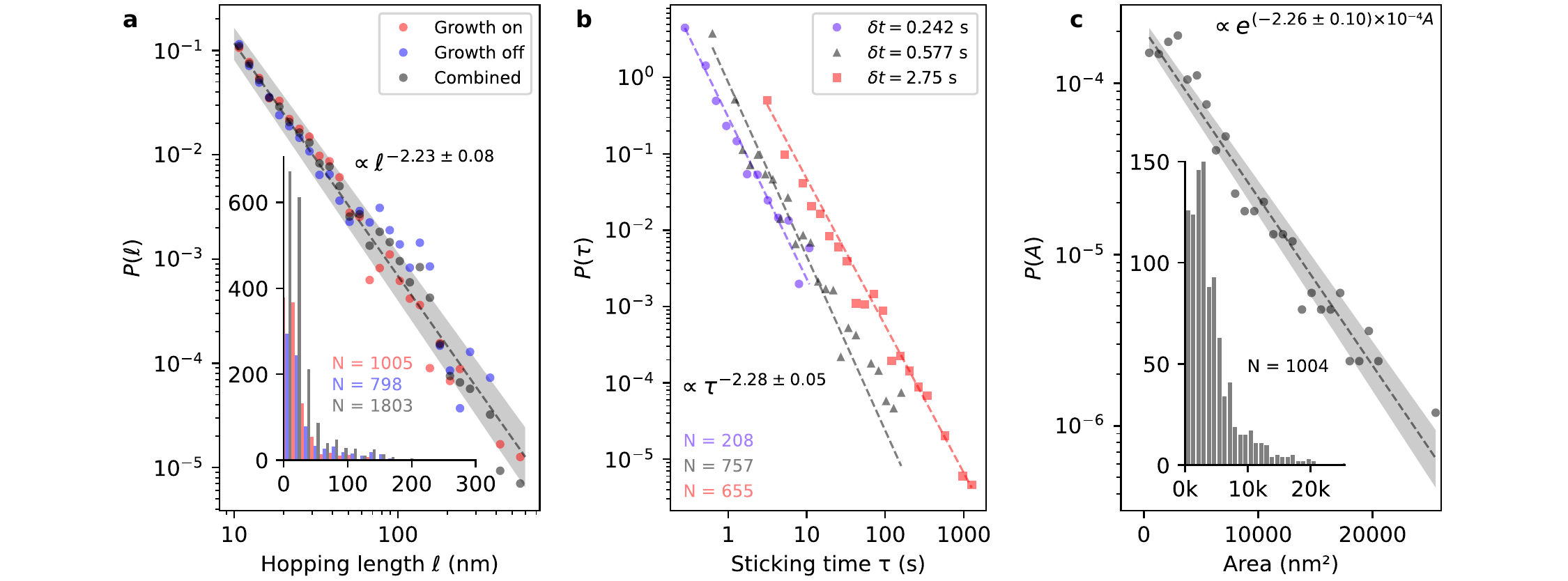}
\caption{\textbf{Statistics of hopping lengths and sticking times.} (a) Log-log histogram of hopping lengths $\ell$ with (red), without (blue) incoming Bi flux and combined (grey). (b) Log-log histogram of sticking times $\tau$ for different frame rates $\delta t = 0.242$~s (blue circles), $0.577$~s (black triangles) and $2.75$~s (red squares). (c) Semi-log histogram of areas during an hopping event. $N$ is the number of events. The shaded areas correspond to the uncertainty in the lines of best fit. The numerical expressions resulting from fitting in (a, b) specified directly in the panels are extracted from the most populated series (`combined' and $\delta t=0.577$~s, respectively). All values of $\eta_\ell$ and $\eta_\tau$ are specified in the text. Insets in (a, c) show the same data in absolute counts on a linear scale.}
\label{fig:3}
\end{figure*}

We now focus on the statistical behaviour of active $\alpha$-Bi islands observed by LEEM by measuring the hopping length $\ell=|\mathbf{r}(t+\delta t)-\mathbf{r}(t)|$ (where $\delta t$ is the time step in LEEM sequences). Figure~\ref{fig:3}(a) shows the distribution $P(\ell)$ with (growth on) and without incoming Bi flux (growth off). The range is limited by pixel size ($\delta x = 10$~nm, hopping events such that $\ell<10$~nm are discarded from analysis). The large majority of diffusion events are characterized with a small hopping length ($80\%$ of the observed hopping events are below 40~nm), while the longest is just under 600~nm. Surprisingly, $P(\ell)$ agrees with a power law distribution $P(\ell)\sim\ell^{-\eta_\ell}$ with $\eta_\ell=2.23\pm0.08$ (see dashed line) independently of the Bi flux ($\eta^\mathrm{on}_\ell=2.34\pm0.09$ and $\eta^\mathrm{off}_\ell=2.12\pm0.11$, best-fit lines hidden for clarity). The distributions $P(\ell)$ of active individual islands are shown in Fig.~\ref{fig:si_allislands} (in SI section \ref{si:statistics}) and evidence the heavy-tailed nature of the hopping length distribution at the individual island level with very similar decay parameters (the majority of islands have $\eta_\ell\simeq2.4$). Note that the decay parameters $\eta$ are determined using maximum likelihood estimation \cite{Clauset2009, Corral2019}.

Besides the hopping lengths, the duration $\tau$ during which an island remains immobilized between successive hopping events, or \emph{sticking time} \cite{Luedtke1999}, is also monitored. The distribution of the sticking times $P(\tau)$ is shown in Fig.~\ref{fig:3}(b). Here, the range is limited by the frame rate $\delta t$ which sets a lower bound on the observable, \emph{i.e.}, $\tau\geq\delta t$. If an island is observed in two distinct locations in successive LEEM frames, then $\tau=\delta t$; if not, $\tau$ is integrated until the next hopping event is observed. $P(\tau)$ also agrees with a power law $\tau^{-\eta_\tau}$ with $\eta_\tau=2.28\pm0.05$ (for the largest population, with $\delta t = 0.577$~s), remarkably similar with the decay parameter in the hopping lengths distributions $P(\ell)$ in Fig.~\ref{fig:3}(a). Distributions obtained for different frame rates ($\delta t = 0.242$~s and $2.75$~s) also show similar values, $\eta_\tau= 2.12\pm0.08$ and $\eta_\tau = 1.93\pm0.04$ respectively.

The areas $A$ for all active islands are extracted at every hopping event during the LEEM sequence. The histogram in Fig.~\ref{fig:3}(c) shows that large islands are less likely to hop, although islands as large as 20000 nm$^2$ show occasional hopping. As opposed to hopping lengths $\ell$ and sticking times $\tau$, the distribution of areas $P(A)$ does not agree with a power-law but instead is described by an exponential distribution $P(A)\sim\exp{(-\alpha\cdot A)}$. This size-dependence is attributed to the fact that larger islands are more likely to be pinned by a substrate defect simply due to the larger contact area at the interface. In fact, we show (see SI section~\ref{si:defects}) that an exponential dependence is in agreement with a model where randomly-distributed point defects in the substrate's surface are responsible for island pinning. The exponential decay parameter $\alpha=(2.26\pm0.10)\times10^{-4}$~nm$^{-2}$ is directly related to the density of graphite point defects in the middle of terraces (excluding the strong pinning of islands decorating terrace step edges); under such model this implies a defect density of $\rho=(2.32\pm0.11)\times10^{10}$~cm$^{-2}$, in rough ballpark with commercial high-quality HOPG \cite{Zhang2015}. Additionally, large islands (at a late stage of deposition with larger coverage) are in closer proximity to their neighbour islands and may coalesce and form grain boundaries \cite{Kowalczyk2012a}, likely to promote pinning and destroy structural superlubricity.


\section{Discussion, outlook and conclusion}

\paragraph{Discussion.} Most random walk processes are governed by exponential distributions, such as Brownian motion \cite{Luedtke1999, Woyczynski2001}. Adsorbed gold clusters and single metallic adatoms on graphene show heavy-tailed hopping distributions in molecular dynamics simulations \cite{Luedtke1999, Gervilla2020} as well as a limited number of experimental systems at liquid-solid interfaces \cite{Chee2016, Chee2019, Wang2020}. Such distributions can be called L{\'e}vy flights when both diffusion lengths $\ell$ and durations $t$ follow power-law distributions with similar exponents $\eta_{t}\simeq\eta_{\ell}>1$. In our experiment the flight durations are unfortunately inaccessible due to the low frame rate in LEEM. Under the hypothesis that the velocities of the islands during a diffusion event are narrowly distributed, it follows that the distribution of hopping durations $P(t)$ should agree with a power law similar to that of $P(\ell)$. Nonetheless, the sticking time distribution $P(\tau)$ in our data agrees with that of a system whose dynamics are identified as L{\'e}vy flights \cite{Luedtke1999}. Previously described in animal colonies \cite{Viswanathan1996, Edwards2007, Humphries2010} (wherein the heavy tail in flight lengths is crucial to the optimization of foraging patterns), these statistics describing random walks are also present in a multitude of other complex systems such as financial \cite{Mantegna1991, Podobnik2011}, geophysical \cite{Corral2006, BeccarVarela2019, Corral2019} and photonic \cite{Barthelemy2008, Kiselev2019}. Experimental realization of L\'evy flights in solid-state diffusion has however remained elusive.

\paragraph{Outlook.}The theory of one-dimensional L{\'e}vy flights in $\alpha$-Bi/graphite is not yet established, although it is clear that the distribution of graphitic superficial defects play a major role, in contrast to \cite{Luedtke1999} where the heavy-tails in the simulated hopping lengths arise without the presence of defects. Despite this, the observed power-law parameters $\eta_\ell$ and $\eta_\tau$ are in excellent agreement with gold clusters/graphite simulations \cite{Luedtke1999}. Future investigations employing \emph{ab-initio} and classical molecular dynamics simulations, in which the role of temperature, defect types and density (as well as morphological parameters such as island size and shape) can be independently investigated, will certainly bring valuable insight. Additionally, nanotribological experiments involving the use of scanning probes which couple force and displacements may further characterize the $\alpha$-Bi/HOPG system in terms of frictional properties.

\paragraph{Conclusion.}Our LEEM experiments evidence spontaneous and directional diffusion of $\alpha$-Bi islands on graphite, explained by a type-B structural superlubricity model identified recently \cite{Panizon2023}. Our RI simulations agree very well with the observations and further evidence a unique directional superlubricity behaviour, where two superlubric pathways (or \emph{nanohighways}) separated by $60^\circ$ are formed when $\alpha$-Bi islands are twisted by $\theta=28^\circ$ and $32^\circ$, which correspond to the observed twist angles in our samples. Finally, statistical analysis of both hopping lengths $\ell$ and sticking times $\tau$ reveal heavy-tailed distributions ($\eta\simeq2.0-2.5$ for both quantities) over many orders of magnitude, indicative of L{\'e}vy flight dynamics in the $\alpha$-Bi/HOPG system. We believe that these results should renew the search for L{\'e}vy flights in solid-state physics and encourage further studies in the tribology of van der Waals contacts with distinct crystalline symmetries.

\section{Methods}

\paragraph{LEEM and LEED.}The experimental data were obtained at the Elettra synchrotron in Trieste, Italy. HOPG (SPI-1) substrates were cleaved in air before loading into the UHV chamber and degassed ($T>400^\circ$C) for about 12 hours. LEEM images were processed and analyzed using ImageJ for tracking; waterfall plots were obtained from cross-sectional profiles across the images. 

\paragraph{Registry index simulations.}Custom-made script based on \cite{Hod2012} was developed for all RI simulations using python.

\paragraph{Statistical analyses.}Statistical analyses were obtained from the tracking data using Python. The decay constants ($\eta_\ell$, $\eta_\tau$ for the hopping lengths and sticking times, respectively) were obtained using maximum likelihood estimation \cite{Clauset2009, Corral2019}, and a standard least-square method (\texttt{scipy.optimize.curve\_fit}) using power law functional for the vertical offsets in log-log plots. The exponential fitting of $P(A)$ was performed using \texttt{scipy.optimize.curve\_fit}.

\section{Additional information}

\paragraph{Author Contributions and Statements.}M.L.S. prepared the manuscript and performed the RI simulations, P.J.K., A.L., T.O.M. and S.A.B. conducted the experiment at the Elettra synchrotron, P.J.K. and K.P. performed the image analysis and statistical analyses. All authors contributed to the discussion equally. The authors declare no competing interests. The data is available upon reasonable request to the corresponding author.

\paragraph{Acknowledgments.}This work was sponsored by the National Science Center, Poland (M.L.S.: 2022/47/D/ST3/03216; P.J.K.: 2018/31/B/ST3/02450) and the MacDiarmid Institute for Advanced Materials and Nanotechnology, New Zealand (S.A.B.). G.B. and T.C.C are supported by the U.S. Department of Energy (DOE), Office of Science, Office of Basic Energy Sciences, Division of Materials Sciences and Engineering, under award number DE-SC0024294 and DE-FG02-07ER46383.

\bibliography{unidirectional}

\begin{thebibliography}{10}
\expandafter\ifx\csname url\endcsname\relax
  \def\url#1{\texttt{#1}}\fi
\expandafter\ifx\csname urlprefix\endcsname\relax\def\urlprefix{URL }\fi
\providecommand{\bibinfo}[2]{#2}
\providecommand{\eprint}[2][]{\url{#2}}

\bibitem{Holmberg2017}
\bibinfo{author}{Holmberg, K.} \& \bibinfo{author}{Erdemir, A.}
\newblock \bibinfo{title}{Influence of tribology on global energy consumption,
  costs and emissions}.
\newblock \emph{\bibinfo{journal}{Friction}} \textbf{\bibinfo{volume}{5}},
  \bibinfo{pages}{263--284} (\bibinfo{year}{2017}).

\bibitem{Luo2021}
\bibinfo{author}{Luo, J.}, \bibinfo{author}{Liu, M.} \& \bibinfo{author}{Ma,
  L.}
\newblock \bibinfo{title}{{Origin of friction and the new frictionless
  technology—Superlubricity: Advancements and future outlook}}.
\newblock \emph{\bibinfo{journal}{Nano Energy}} \textbf{\bibinfo{volume}{86}},
  \bibinfo{pages}{106092} (\bibinfo{year}{2021}).

\bibitem{Hod2012}
\bibinfo{author}{Hod, O.}
\newblock \bibinfo{title}{Interlayer commensurability and superlubricity in
  rigid layered materials}.
\newblock \emph{\bibinfo{journal}{Phys. Rev. B}} \textbf{\bibinfo{volume}{86}},
  \bibinfo{pages}{075444} (\bibinfo{year}{2012}).

\bibitem{Hod2018}
\bibinfo{author}{Hod, O.}, \bibinfo{author}{Meyer, E.}, \bibinfo{author}{Zheng,
  Q.} \& \bibinfo{author}{Urbakh, M.}
\newblock \bibinfo{title}{Structural superlubricity and ultralow friction
  across the length scales}.
\newblock \emph{\bibinfo{journal}{Nature}} \textbf{\bibinfo{volume}{563}},
  \bibinfo{pages}{485--492} (\bibinfo{year}{2018}).

\bibitem{Martin2018}
\bibinfo{author}{Martin, J.~M.} \& \bibinfo{author}{Erdemir, A.}
\newblock \bibinfo{title}{{Superlubricity: Friction's vanishing act}}.
\newblock \emph{\bibinfo{journal}{Phys. Today}} \textbf{\bibinfo{volume}{71}},
  \bibinfo{pages}{40--46} (\bibinfo{year}{2018}).

\bibitem{Zhai2019}
\bibinfo{author}{Zhai, W.}, \bibinfo{author}{Zhai, W.} \&
  \bibinfo{author}{Zhou, K.}
\newblock \bibinfo{title}{{Nanomaterials in Superlubricity}}.
\newblock \emph{\bibinfo{journal}{Adv. Funct. Mater.}}
  \textbf{\bibinfo{volume}{29}}, \bibinfo{pages}{1806395}
  (\bibinfo{year}{2019}).

\bibitem{Vanossi2020}
\bibinfo{author}{Vanossi, A.}, \bibinfo{author}{Vanossi, A.},
  \bibinfo{author}{Bechinger, C.} \& \bibinfo{author}{Urbakh, M.}
\newblock \bibinfo{title}{Structural lubricity in soft and hard matter
  systems}.
\newblock \emph{\bibinfo{journal}{Nat. Commun.}} \textbf{\bibinfo{volume}{11}},
  \bibinfo{pages}{4657} (\bibinfo{year}{2020}).

\bibitem{Wang2024}
\bibinfo{author}{Wang, J.}, \bibinfo{author}{Khosravi, A.},
  \bibinfo{author}{Vanossi, A.} \& \bibinfo{author}{Tosatti, E.}
\newblock \bibinfo{title}{{Colloquium: Sliding and pinning in structurally
  lubric 2D material interfaces}}.
\newblock \emph{\bibinfo{journal}{Rev. Mod. Phys.}}
  \textbf{\bibinfo{volume}{96}}, \bibinfo{pages}{011002}
  (\bibinfo{year}{2024}).

\bibitem{Ouyang2018}
\bibinfo{author}{Ouyang, W.}, \bibinfo{author}{Mandelli, D.},
  \bibinfo{author}{Urbakh, M.} \& \bibinfo{author}{Hod, O.}
\newblock \bibinfo{title}{{Nanoserpents: Graphene Nanoribbon Motion on
  Two-Dimensional Hexagonal Materials}}.
\newblock \emph{\bibinfo{journal}{Nano Lett.}} \textbf{\bibinfo{volume}{18}},
  \bibinfo{pages}{6009--6016} (\bibinfo{year}{2018}).

\bibitem{Androulidakis2020}
\bibinfo{author}{Androulidakis, C.}, \bibinfo{author}{Koukaras, E.~N.},
  \bibinfo{author}{Paterakis, G.}, \bibinfo{author}{Trakakis, G.} \&
  \bibinfo{author}{Galiotis, C.}
\newblock \bibinfo{title}{Tunable macroscale structural superlubricity in
  two-layer graphene via strain engineering}.
\newblock \emph{\bibinfo{journal}{Nat. Commun.}} \textbf{\bibinfo{volume}{11}},
  \bibinfo{pages}{1595} (\bibinfo{year}{2020}).

\bibitem{Liao2021}
\bibinfo{author}{Liao, M.} \emph{et~al.}
\newblock \bibinfo{title}{{Ultra-low friction and edge-pinning effect in
  large-lattice-mismatch van der Waals heterostructures}}.
\newblock \emph{\bibinfo{journal}{Nat. Mater.}} \textbf{\bibinfo{volume}{21}},
  \bibinfo{pages}{47--53} (\bibinfo{year}{2021}).

\bibitem{Losi2020}
\bibinfo{author}{Losi, G.}, \bibinfo{author}{Restuccia, P.} \&
  \bibinfo{author}{Righi, M.~C.}
\newblock \bibinfo{title}{Superlubricity in phosphorene identified by means of
  ab initio calculations}.
\newblock \emph{\bibinfo{journal}{2D Mater.}} \textbf{\bibinfo{volume}{7}},
  \bibinfo{pages}{025033} (\bibinfo{year}{2020}).

\bibitem{Panizon2023}
\bibinfo{author}{Panizon, E.} \emph{et~al.}
\newblock \bibinfo{title}{{Frictionless nanohighways on crystalline surfaces}}.
\newblock \emph{\bibinfo{journal}{Nanoscale}} \textbf{\bibinfo{volume}{15}},
  \bibinfo{pages}{1299--1316} (\bibinfo{year}{2023}).

\bibitem{Trillitzsch2018}
\bibinfo{author}{Trillitzsch, F.} \emph{et~al.}
\newblock \bibinfo{title}{{Directional and angular locking in the driven motion
  of Au islands on MoS$\ensuremath{_2}$}}.
\newblock \emph{\bibinfo{journal}{Phys. Rev. B}} \textbf{\bibinfo{volume}{98}},
  \bibinfo{pages}{165417} (\bibinfo{year}{2018}).

\bibitem{Cao2021}
\bibinfo{author}{Cao, X.} \emph{et~al.}
\newblock \bibinfo{title}{Pervasive orientational and directional locking at
  geometrically heterogeneous sliding interfaces}.
\newblock \emph{\bibinfo{journal}{Phys. Rev. E}}
  \textbf{\bibinfo{volume}{103}}, \bibinfo{pages}{012606}
  (\bibinfo{year}{2021}).

\bibitem{Scott2005}
\bibinfo{author}{Scott, S.~A.}, \bibinfo{author}{Kral, M.~V.} \&
  \bibinfo{author}{Brown, S.~A.}
\newblock \bibinfo{title}{{Growth of oriented {Bi} nanorods at graphite
  step-edges}}.
\newblock \emph{\bibinfo{journal}{Phys. Rev. B}} \textbf{\bibinfo{volume}{72}},
  \bibinfo{pages}{205423--1--8} (\bibinfo{year}{2005}).

\bibitem{McCarthy2010}
\bibinfo{author}{McCarthy, D.~N.}, \bibinfo{author}{Robertson, D.},
  \bibinfo{author}{Kowalczyk, P.~J.} \& \bibinfo{author}{Brown, S.~A.}
\newblock \bibinfo{title}{{The effects of annealing and growth temperature on
  the morphologies of {Bi} nanostructures on {HOPG}}}.
\newblock \emph{\bibinfo{journal}{Surf. Sci.}} \textbf{\bibinfo{volume}{604}},
  \bibinfo{pages}{1273--1282} (\bibinfo{year}{2010}).

\bibitem{Kowalczyk2011}
\bibinfo{author}{Kowalczyk, P.~J.} \emph{et~al.}
\newblock \bibinfo{title}{{{STM} and {XPS} investigations of bismuth islands on
  {HOPG}}}.
\newblock \emph{\bibinfo{journal}{Surf. Sci.}} \textbf{\bibinfo{volume}{605}},
  \bibinfo{pages}{659 -- 667} (\bibinfo{year}{2011}).

\bibitem{Kowalczyk2013}
\bibinfo{author}{Kowalczyk, P.~J.} \emph{et~al.}
\newblock \bibinfo{title}{{Electronic Size Effects in Three-Dimensional
  Nanostructures}}.
\newblock \emph{\bibinfo{journal}{Nano Lett.}} \textbf{\bibinfo{volume}{13}},
  \bibinfo{pages}{43--47} (\bibinfo{year}{2013}).

\bibitem{LeSter2019}
\bibinfo{author}{Le~Ster, M.}, \bibinfo{author}{Maerkl, T.},
  \bibinfo{author}{Kowalczyk, P.~J.} \& \bibinfo{author}{Brown, S.~A.}
\newblock \bibinfo{title}{{Moir{\'{e}} patterns in van der Waals
  heterostructures}}.
\newblock \emph{\bibinfo{journal}{Phys. Rev. B}} \textbf{\bibinfo{volume}{99}},
  \bibinfo{pages}{075422} (\bibinfo{year}{2019}).

\bibitem{Kowalczyk2020}
\bibinfo{author}{Kowalczyk, P.~J.} \emph{et~al.}
\newblock \bibinfo{title}{{Realization of Symmetry Enforced Two-Dimensional
  Dirac Fermions in Nonsymmorphic $\alpha$-Bismuthene}}.
\newblock \emph{\bibinfo{journal}{{ACS} Nano}} \textbf{\bibinfo{volume}{14}},
  \bibinfo{pages}{1888--1894} (\bibinfo{year}{2020}).

\bibitem{Salehi2023}
\bibinfo{author}{Salehitaleghani, S.} \emph{et~al.}
\newblock \bibinfo{title}{{Edge states of $\ensuremath{\alpha}$-bismuthene
  nanostructures}}.
\newblock \emph{\bibinfo{journal}{2D Mater.}} \textbf{\bibinfo{volume}{10}},
  \bibinfo{pages}{015020} (\bibinfo{year}{2022}).

\bibitem{Locatelli2006}
\bibinfo{author}{Locatelli, A.}, \bibinfo{author}{Aballe, L.},
  \bibinfo{author}{Mentes, T.~O.}, \bibinfo{author}{Kiskinova, M.} \&
  \bibinfo{author}{Bauer, E.}
\newblock \bibinfo{title}{{Photoemission electron microscopy with chemical
  sensitivity: SPELEEM methods and applications}}.
\newblock \emph{\bibinfo{journal}{Surf. Inter. Anal.}}
  \textbf{\bibinfo{volume}{38}}, \bibinfo{pages}{1554--1557}
  (\bibinfo{year}{2006}).

\bibitem{Mentes2012}
\bibinfo{author}{Mente\c{s}, T.~O.} \& \bibinfo{author}{Locatelli, A.}
\newblock \bibinfo{title}{{Angle-resolved X-ray photoemission electron
  microscopy}}.
\newblock \emph{\bibinfo{journal}{J. Electron Spectrosc. Relat. Phenom.}}
  \textbf{\bibinfo{volume}{185}}, \bibinfo{pages}{323 -- 329}
  (\bibinfo{year}{2012}).

\bibitem{Mentes2014}
\bibinfo{author}{Mente\c{s}, T.~O.}, \bibinfo{author}{Zamborlini, G.},
  \bibinfo{author}{Sala, A.} \& \bibinfo{author}{Locatelli, A.}
\newblock \bibinfo{title}{Cathode lens spectromicroscopy: methodology and
  applications}.
\newblock \emph{\bibinfo{journal}{Beilstein J. of Nanotechnol.}}
  \textbf{\bibinfo{volume}{5}}, \bibinfo{pages}{1873--1886}
  (\bibinfo{year}{2014}).

\bibitem{Kowalczyk2015}
\bibinfo{author}{Kowalczyk, P.~J.} \emph{et~al.}
\newblock \bibinfo{title}{{Origin of the moir\'e pattern in thin {Bi} films
  deposited on {HOPG}}}.
\newblock \emph{\bibinfo{journal}{Phys. Rev. B}} \textbf{\bibinfo{volume}{91}},
  \bibinfo{pages}{045434} (\bibinfo{year}{2015}).

\bibitem{Kowalczyk2014}
\bibinfo{author}{Kowalczyk, P.} \emph{et~al.}
\newblock \bibinfo{title}{{STM driven modification of bismuth nanostructures}}.
\newblock \emph{\bibinfo{journal}{Surf. Sci.}} \textbf{\bibinfo{volume}{621}},
  \bibinfo{pages}{140--145} (\bibinfo{year}{2014}).

\bibitem{Chee2019}
\bibinfo{author}{Chee, S.~W.}, \bibinfo{author}{Anand, U.},
  \bibinfo{author}{Bisht, G.}, \bibinfo{author}{Tan, S.~F.} \&
  \bibinfo{author}{Mirsaidov, U.}
\newblock \bibinfo{title}{{Direct Observations of the Rotation and Translation
  of Anisotropic Nanoparticles Adsorbed at a Liquid{\textendash}Solid
  Interface}}.
\newblock \emph{\bibinfo{journal}{Nano Lett.}} \textbf{\bibinfo{volume}{19}},
  \bibinfo{pages}{2871--2878} (\bibinfo{year}{2019}).

\bibitem{Yesibolati2020}
\bibinfo{author}{Yesibolati, M.~N.} \emph{et~al.}
\newblock \bibinfo{title}{{Unhindered Brownian Motion of Individual
  Nanoparticles in Liquid-Phase Scanning Transmission Electron Microscopy}}.
\newblock \emph{\bibinfo{journal}{Nano Lett.}} \textbf{\bibinfo{volume}{20}},
  \bibinfo{pages}{7108--7115} (\bibinfo{year}{2020}).

\bibitem{Jia2020}
\bibinfo{author}{Jia, H.}, \bibinfo{author}{Guo, Q.} \& \bibinfo{author}{Du,
  S.}
\newblock \bibinfo{title}{{Thermally Driven Diffusion of a Magic Number
  Gold{\textendash}Fullerene Cluster on a Au(111) Surface}}.
\newblock \emph{\bibinfo{journal}{J. Phys. Chem. C}}
  \textbf{\bibinfo{volume}{124}}, \bibinfo{pages}{9990--9995}
  (\bibinfo{year}{2020}).

\bibitem{Vasileiadis2019}
\bibinfo{author}{Vasileiadis, T.} \emph{et~al.}
\newblock \bibinfo{title}{{Ultrafast rotational motions of supported
  nanoclusters probed by electron diffraction}}.
\newblock \emph{\bibinfo{journal}{Nanoscale Horiz.}}
  \textbf{\bibinfo{volume}{4}}, \bibinfo{pages}{1164--1173}
  (\bibinfo{year}{2019}).

\bibitem{Clauset2009}
\bibinfo{author}{Clauset, A.}, \bibinfo{author}{Shalizi, C.~R.} \&
  \bibinfo{author}{Newman, M. E.~J.}
\newblock \bibinfo{title}{{Power-Law Distributions in Empirical Data}}.
\newblock \emph{\bibinfo{journal}{{SIAM} Review}}
  \textbf{\bibinfo{volume}{51}}, \bibinfo{pages}{661--703}
  (\bibinfo{year}{2009}).

\bibitem{Corral2019}
\bibinfo{author}{Corral, {\'{A}}.} \& \bibinfo{author}{Gonz{\'{a}}lez,
  {\'{A}}.}
\newblock \bibinfo{title}{{Power Law Size Distributions in Geoscience
  Revisited}}.
\newblock \emph{\bibinfo{journal}{Earth Space Sci.}}
  \textbf{\bibinfo{volume}{6}}, \bibinfo{pages}{673} (\bibinfo{year}{2019}).

\bibitem{Luedtke1999}
\bibinfo{author}{Luedtke, W.~D.} \& \bibinfo{author}{Landman, U.}
\newblock \bibinfo{title}{{Slip Diffusion and L{\'{e}}vy Flights of an Adsorbed
  Gold Nanocluster}}.
\newblock \emph{\bibinfo{journal}{Phys. Rev. Lett.}}
  \textbf{\bibinfo{volume}{82}}, \bibinfo{pages}{3835--3838}
  (\bibinfo{year}{1999}).

\bibitem{Zhang2015}
\bibinfo{author}{Zhang, G.}, \bibinfo{author}{Cuharuc, A.~S.},
  \bibinfo{author}{Güell, A.~G.} \& \bibinfo{author}{Unwin, P.~R.}
\newblock \bibinfo{title}{{Electrochemistry at highly oriented pyrolytic
  graphite (HOPG): lower limit for the kinetics of outer-sphere redox processes
  and general implications for electron transfer models}}.
\newblock \emph{\bibinfo{journal}{Phys. Chem. Chem. Phys.}}
  \textbf{\bibinfo{volume}{17}}, \bibinfo{pages}{11827--11838}
  (\bibinfo{year}{2015}).

\bibitem{Kowalczyk2012a}
\bibinfo{author}{Kowalczyk, P.~J.}, \bibinfo{author}{Beli\'{c}, D.},
  \bibinfo{author}{Mahapatra, O.} \& \bibinfo{author}{Brown, S.~A.}
\newblock \bibinfo{title}{Grain boundaries between bismuth nanocrystals}.
\newblock \emph{\bibinfo{journal}{Acta Mater.}} \textbf{\bibinfo{volume}{60}},
  \bibinfo{pages}{674 -- 681} (\bibinfo{year}{2012}).

\bibitem{Woyczynski2001}
\bibinfo{author}{Woyczy{\'{n}}ski, W.~A.}
\newblock \emph{\bibinfo{title}{L{\'e}vy Processes in the Physical Sciences}},
  \bibinfo{pages}{241--266} (\bibinfo{publisher}{Birkh{\"a}user Boston},
  \bibinfo{address}{Boston, MA}, \bibinfo{year}{2001}).

\bibitem{Gervilla2020}
\bibinfo{author}{Gervilla, V.}, \bibinfo{author}{Zarshenas, M.},
  \bibinfo{author}{Sangiovanni, D.~G.} \& \bibinfo{author}{Sarakinos, K.}
\newblock \bibinfo{title}{{Anomalous vs. Normal Room-temperature Diffusion of
  Metal Adatoms on Graphene}}.
\newblock \emph{\bibinfo{journal}{J. Phys. Chem. C}}
  \textbf{\bibinfo{volume}{11}}, \bibinfo{pages}{8930--8936}
  (\bibinfo{year}{2020}).

\bibitem{Chee2016}
\bibinfo{author}{Chee, S.~W.}, \bibinfo{author}{Baraissov, Z.},
  \bibinfo{author}{Loh, N.~D.}, \bibinfo{author}{Matsudaira, P.~T.} \&
  \bibinfo{author}{Mirsaidov, U.}
\newblock \bibinfo{title}{{Desorption-Mediated Motion of Nanoparticles at the
  Liquid{\textendash}Solid Interface}}.
\newblock \emph{\bibinfo{journal}{J. Phys. Chem. C}}
  \textbf{\bibinfo{volume}{120}}, \bibinfo{pages}{20462--20470}
  (\bibinfo{year}{2016}).

\bibitem{Wang2020}
\bibinfo{author}{Wang, D.} \& \bibinfo{author}{Schwartz, D.~K.}
\newblock \bibinfo{title}{{Non-Brownian Interfacial Diffusion: Flying, Hopping,
  and Crawling}}.
\newblock \emph{\bibinfo{journal}{J. Phys. Chem. C}}
  \textbf{\bibinfo{volume}{124}}, \bibinfo{pages}{19880--19891}
  (\bibinfo{year}{2020}).

\bibitem{Viswanathan1996}
\bibinfo{author}{Viswanathan, G.~M.} \emph{et~al.}
\newblock \bibinfo{title}{{L{\'{e}}vy flight search patterns of wandering
  albatrosses}}.
\newblock \emph{\bibinfo{journal}{Nature}} \textbf{\bibinfo{volume}{381}},
  \bibinfo{pages}{413--415} (\bibinfo{year}{1996}).

\bibitem{Edwards2007}
\bibinfo{author}{Edwards, A.~M.} \emph{et~al.}
\newblock \bibinfo{title}{{Revisiting L{\'{e}}vy flight search patterns of
  wandering albatrosses, bumblebees and deer}}.
\newblock \emph{\bibinfo{journal}{Nature}} \textbf{\bibinfo{volume}{449}},
  \bibinfo{pages}{1044--1048} (\bibinfo{year}{2007}).

\bibitem{Humphries2010}
\bibinfo{author}{Humphries, N.~E.} \emph{et~al.}
\newblock \bibinfo{title}{{Environmental context explains L{\'{e}}vy and
  Brownian movement patterns of marine predators}}.
\newblock \emph{\bibinfo{journal}{Nature}} \textbf{\bibinfo{volume}{465}},
  \bibinfo{pages}{1066--1069} (\bibinfo{year}{2010}).

\bibitem{Mantegna1991}
\bibinfo{author}{Mantegna, R.~N.}
\newblock \bibinfo{title}{{L{\'{e}}vy walks and enhanced diffusion in Milan
  stock exchange}}.
\newblock \emph{\bibinfo{journal}{Phys. A: Stat.}}
  \textbf{\bibinfo{volume}{179}}, \bibinfo{pages}{232--242}
  (\bibinfo{year}{1991}).

\bibitem{Podobnik2011}
\bibinfo{author}{Podobnik, B.}, \bibinfo{author}{Valentincic, A.},
  \bibinfo{author}{Horvatic, D.} \& \bibinfo{author}{Stanley, H.~E.}
\newblock \bibinfo{title}{{Asymmetric Levy flight in financial ratios}}.
\newblock \emph{\bibinfo{journal}{PNAS}} \textbf{\bibinfo{volume}{108}},
  \bibinfo{pages}{17883--17888} (\bibinfo{year}{2011}).

\bibitem{Corral2006}
\bibinfo{author}{Corral, {\'{A}}.}
\newblock \bibinfo{title}{{Universal Earthquake-Occurrence Jumps, Correlations
  with Time, and Anomalous Diffusion}}.
\newblock \emph{\bibinfo{journal}{Phys. Rev. Lett.}}
  \textbf{\bibinfo{volume}{97}}, \bibinfo{pages}{178501}
  (\bibinfo{year}{2006}).

\bibitem{BeccarVarela2019}
\bibinfo{author}{Beccar-Varela, M.~P.}, \bibinfo{author}{Gonzalez-Huizar, H.},
  \bibinfo{author}{Mariani, M.~C.} \& \bibinfo{author}{Tweneboah, O.~K.}
\newblock \bibinfo{title}{{L{\'{e}}vy Flights and Wavelets Analysis of
  Volcano-Seismic Data}}.
\newblock \emph{\bibinfo{journal}{Pure Appl. Geophys.}}
  \textbf{\bibinfo{volume}{177}}, \bibinfo{pages}{723--736}
  (\bibinfo{year}{2019}).

\bibitem{Barthelemy2008}
\bibinfo{author}{Barthelemy, P.}, \bibinfo{author}{Bertolotti, J.} \&
  \bibinfo{author}{Wiersma, D.~S.}
\newblock \bibinfo{title}{{A L{\'{e}}vy flight for light}}.
\newblock \emph{\bibinfo{journal}{Nature}} \textbf{\bibinfo{volume}{453}},
  \bibinfo{pages}{495--498} (\bibinfo{year}{2008}).

\bibitem{Kiselev2019}
\bibinfo{author}{Kiselev, E.~I.} \& \bibinfo{author}{Schmalian, J.}
\newblock \bibinfo{title}{{L{\'{e}}vy Flights and Hydrodynamic Superdiffusion
  on the Dirac Cone of Graphene}}.
\newblock \emph{\bibinfo{journal}{Phys. Rev. Lett.}}
  \textbf{\bibinfo{volume}{123}}, \bibinfo{pages}{195302}
  (\bibinfo{year}{2019}).

\bibitem{Takahashi2022}
\bibinfo{author}{Takahashi, K.}, \bibinfo{author}{Imamura, M.},
  \bibinfo{author}{Yamamoto, I.} \& \bibinfo{author}{Azuma, J.}
\newblock \bibinfo{title}{{Thickness dependent band structure of
  $\ensuremath{\alpha}$-bismuthene grown on epitaxial graphene}}.
\newblock \emph{\bibinfo{journal}{J. Phys. Condens. Matter}}
  \textbf{\bibinfo{volume}{34}}, \bibinfo{pages}{235502}
  (\bibinfo{year}{2022}).

\bibitem{Bai2022}
\bibinfo{author}{Bai, Y.} \emph{et~al.}
\newblock \bibinfo{title}{{Doubled quantum spin Hall effect with high-spin
  Chern number in $\ensuremath{\alpha}$-antimonene and
  $\ensuremath{\alpha}$-bismuthene}}.
\newblock \emph{\bibinfo{journal}{Phys. Rev. B}}
  \textbf{\bibinfo{volume}{105}}, \bibinfo{pages}{195142}
  (\bibinfo{year}{2022}).

\end{thebibliography}

\clearpage
\newpage
\setcounter{section}{0}
\setcounter{figure}{0}
\setcounter{table}{0}
\renewcommand\thesection{S\arabic{section}}
\renewcommand\thefigure{S\arabic{figure}}
\renewcommand\thetable{S\arabic{table}}

{\centering \large \textbf{Supplementary Information}}
\vspace{2mm}
\hrule
\vspace{2mm}
{\centering Evidence of directional structural superlubricity and L{\'e}vy flights in a van der Waals heterostructure}
\vspace{2mm}
\hrule
\vspace{2mm}
{\centering M. Le Ster \emph{et al.}}


\section{Coincidence Vectors}
\label{si:coincidencevectors}

Figure~\ref{fig:si1}(a) shows the reciprocal lattices of $\alpha$-Bi (red) and graphite (black) for $\theta=30^\circ$, similar to the observed values of the twist angles. It is clear that $\mathbf{Bi}(\bar{1}2)$ and $\mathbf{G}(\bar{1}1)$ and $\mathbf{Bi}(12)$ and $\mathbf{G}(01)$ are in close proximity in reciprocal space and are therefore (independently) candidates for type-B commensurability \cite{Panizon2023}. We define the two lattices such that either $\mathbf{Bi}(\bar{12})$ and $\mathbf{G}(\bar{1}1)$, or $\mathbf{Bi}(\bar{1}2)$ and $\mathbf{G}(01)$ are superposed. The two distinct coincidence conditions are:
\begin{figure}[t!]
\centering
\includegraphics[width=\columnwidth]{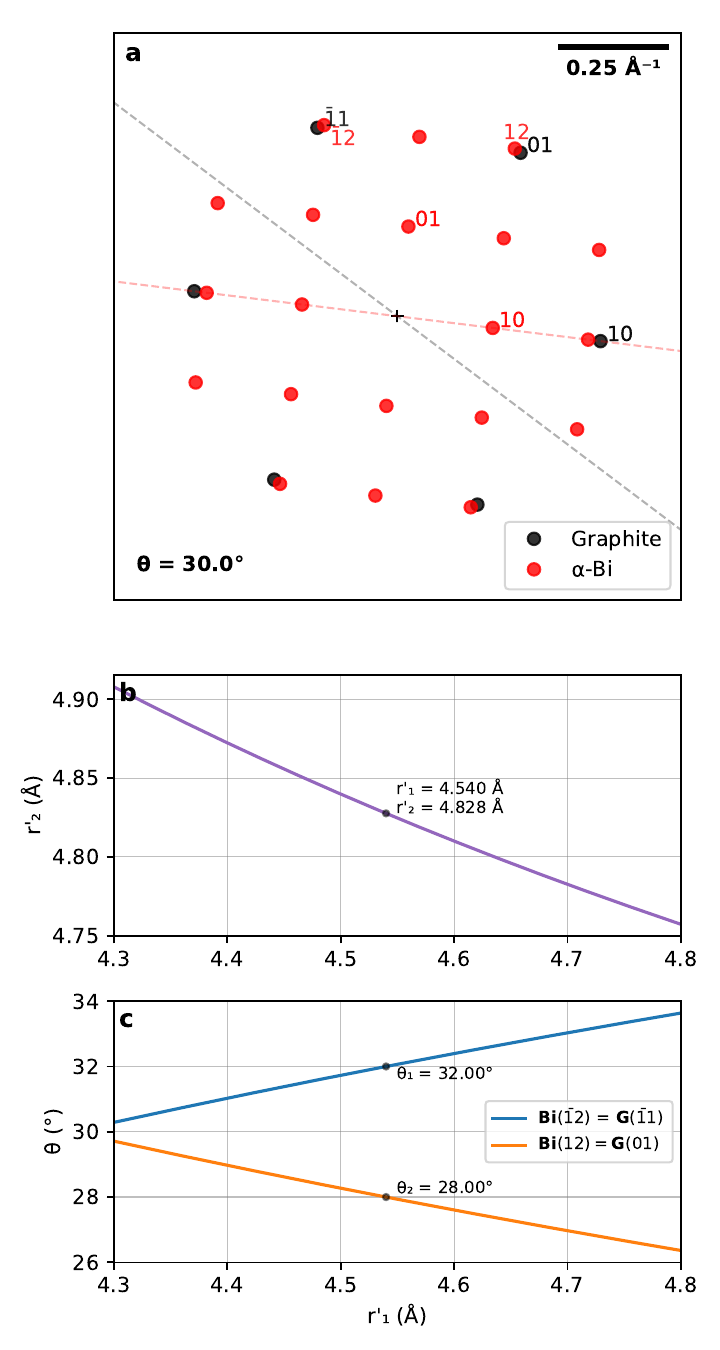}
\caption{\textbf{One-dimensional commensurate condition in reciprocal space.} (a) Reciprocal lattices of $\alpha$-Bi (red) and graphite (black) for $\theta=30^\circ$. The dashed lines correspond to the zigzag directions of both crystals (separated by $\theta$). The reciprocal lattice points pairs ($\mathbf{Bi}(\bar{1}2)$, $\mathbf{G}(\bar{1}1)$) and ($\mathbf{Bi}(12)$, $\mathbf{G}(01)$) are in close proximity. (b) $\alpha$-Bi's lattice parameters $(r_1, r_2)$ satisfying both $\mathbf{Bi}(\bar{1}2) = \mathbf{G}(\bar{1}1)$ and $\mathbf{Bi}(12) = \mathbf{G}(01)$, which are equivalent by symmetry. (c) Twist angle $\theta_1$ and $\theta_2$ satisfying both commensurate definitions as a function of $r'_1$.}
\label{fig:si1}
\end{figure}
\begin{equation}
\mathbf{Bi}(\bar{1}2)(\theta_1) = \mathbf{G}(\bar{1}1)
\label{eq:case1}
\end{equation}
\begin{equation}
\mathbf{Bi}(12)(\theta_2) = \mathbf{G}(01).
\label{eq:case2}
\end{equation}
We develop eqs.~(\ref{eq:case1}, \ref{eq:case2}) to obtain $\theta_1$ and $\theta_2$. With the 2D unit cells defined as in our previous work \cite{LeSter2019} with $\mathbf{R}_1, \mathbf{R}_2$ vectors (with lengths $r_1$, $r_2$, respectively) spanning the real space unit cell and separated by an angle $\omega$ ($\omega$ referring to the substrate, here graphite's $\omega=120^\circ$, and $\alpha$-Bi's $\omega'=90^\circ$). Conditions~(\ref{eq:case1}) and (\ref{eq:case2}) imply a lattice parameter condition (here defined in reciprocal space) as follows:
\begin{equation}
\label{eq:latticeparams}
\begin{cases}
&||\mathbf{Bi}(\bar{1}2)|| = ||\mathbf{G}(\bar{1}1)||\\
&||\mathbf{Bi}(12)|| = ||\mathbf{G}(01)||.
\end{cases}
\end{equation}
By symmetry, the two equations in eq. (\ref{eq:latticeparams}) are identical. Either can be independently developed into the condition dictating the real space lattice parameters $r'_1$ and $r'_2$ of $\alpha$-Bi as follows:
\begin{equation}
\label{eq:r2prime}
r'_2 = \frac{1}{\sqrt{\frac{1}{3a^2}-\frac{1}{4r'^2_1}}}
\end{equation}
valid for $r'_1>\frac{a\sqrt{3}}{2}=2.13...$~($r'_1\simeq4.5$~\AA, so eq.~(\ref{eq:r2prime}) has a valid domain of applicability). Figure~\ref{fig:si1}(b) shows the set of lattice parameters $(r'_1, r'_2)$ such that $\alpha$-Bi/graphite is commensurate (for both conditions in eqs.~(\ref{eq:case1},~\ref{eq:case2})) using $a=2.461$~\AA. Additionally, a twist angle condition imposed by commensurability conditions in eqs.~(\ref{eq:case1},~\ref{eq:case2}) must be considered, which are determined by basic trigonometry. The resulting twist angles $\theta$ as a function of $r'_1$ are shown in Fig.~\ref{fig:si1}(c). The two twist values $\theta_1 = 32.00^\circ$ and $\theta_2 = 28.00^\circ$ are in very good agreement with the experimental observations in the main paper. Bismuthene's lattice parameters can be modified slightly whilst maintaining the commensurate conditions~(\ref{eq:case1}) or~(\ref{eq:case2}) as shown in Fig.~\ref{fig:si1}(b, c).

Two other reciprocal lattice points are in relative proximity: $\mathbf{Bi}(20)$ and $\mathbf{G}(10)$. Under sufficient uniaxial compressive strain of the $\alpha$-Bi layer ($\varepsilon_{1} = -6.1\%$) the 1D commensurate condition can be met for $\theta=30^\circ$, which is likely to imply $\mathbf{Bi}(12)=\mathbf{G}(01)$ and $\mathbf{Bi}(\bar{1}2)=\mathbf{G}(\bar{1}1)$, in this case satisfied with an additional tensile strain along the perpendicular direction $\varepsilon_2 = +1.9\%$. Such 2D commensurate condition does not correspond to the lattice parameters of $\alpha$-Bi on graphite and disagrees with our $\mu$-LEED experiments, and would correspond to a standard commensurate type-A contact, prohibiting superlubricity. The presence of an easy hopping direction in the experiments allows to discard such epitaxial relationship.

To summarize, there are two commensurate cases for $\alpha$-Bi/graphite in the vicinity of $\theta=30^\circ$ for a fixed rigid lattice ($4.540\times4.828$~\AA$^2$) minimizing strain with respect to accepted lattice constant values \cite{Kowalczyk2013, Kowalczyk2015, LeSter2019, Kowalczyk2020, Takahashi2022, Bai2022}, \emph{i.e.}, for $\theta_1=32^\circ$ and $\theta_2=28^\circ$ which correspond to $\mathbf{Bi}(\bar{1}2)=\mathbf{G}(\bar{1}1)$ and $\mathbf{Bi}(12)=\mathbf{G}(01)$, respectively. A single pair of reciprocal lattice points from bismuthene and graphite can be superimposed at once ($\Omega = \{\mathbf{Bi}(12)\}$ or $\Omega = \{\mathbf{Bi}(\bar{1}2)\}$), indicative of a type-B contact \cite{Panizon2023}.


\section{Diffusion velocity}
\label{si:velocity}

\begin{figure}[t]
\centering
\includegraphics[width=\columnwidth]{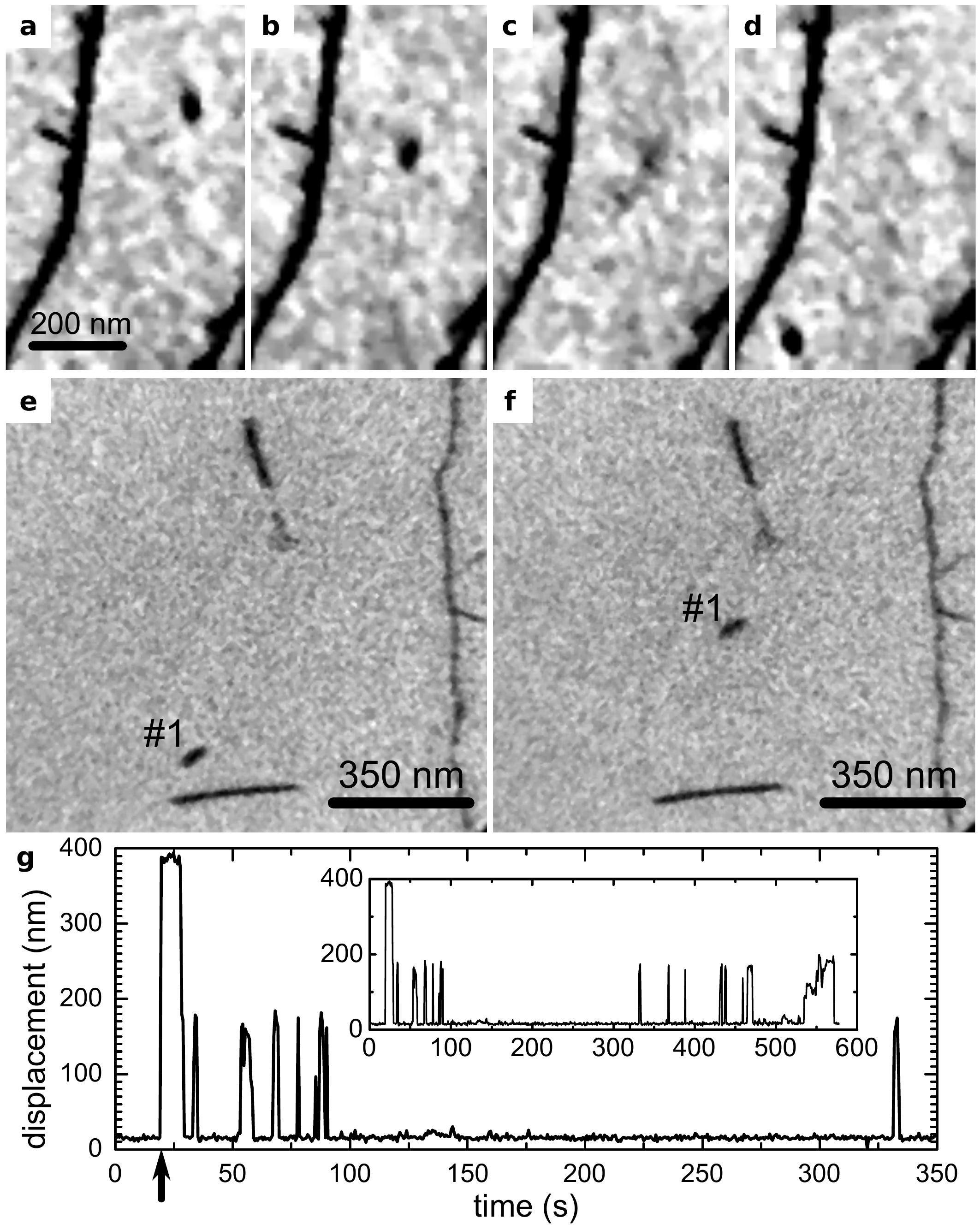}
\caption{\textbf{Estimating the diffusion velocity.} (a-d) Four consecutive LEEM images showing the diffusion of a single $\alpha$-Bi island. (e, f) Two consecutive LEEM frames (in a different sequence) showing rapid movement of the island labelled \#1. (g) Trajectory of island \#1 extracted for a 350~s duration (inset: trajectory over a 600~s period). The arrow indicates the time at which frames (e) and (f) are obtained. See MOV-2 for the LEEM sequence corresponding to (e-g).}
\label{fig:si:velocity}
\end{figure}

The LEEM experiments allow an estimate of the diffusion velocity of $\alpha$-Bi islands on HOPG. Three approaches to estimate the diffusion velocity $v_d$ are explained below.

\paragraph{Case 1.}Island 3 in the main text in Fig.~1(g) is captured in two distinct locations. This indicates that the diffusion velocity is much faster than the velocity limit imposed by the LEEM frame rate ($\delta t = 2.75$~s). The LEEM image shows that the island has to be in location $3$ and $3'$ within the shorter span of the acquisition time ($\delta t_\mathrm{acq} = 350$~ms), during which the island is recorded in the two separate positions ($\Delta x \simeq 100$~nm) for 30-40\% of the acquisition time, \emph{i.e.}, 140~ms. This allows for an estimate of the lower bound of the diffusion velocity, approximately $v_d = 700$~nm/s.

\paragraph{Case 2.}Fig.~\ref{fig:si:velocity} shows four consecutive LEEM images showing the diffusion of a single $\alpha$-Bi island. The island diffuses from the top of the image in Fig.~\ref{fig:si:velocity}(a) by about 100~nm lower in the next LEEM image in Fig.~\ref{fig:si:velocity}(b). The island in Fig.~\ref{fig:si:velocity}(c) is smeared indicating that it was captured during a diffusion event. Finally, Fig.~\ref{fig:si:velocity}(d) shows the island in its final position, approximately 550~nm away from the initial position in Fig.~\ref{fig:si:velocity}(a). The smearing distance ($\Delta x \simeq 150$~nm) and the acquisition time ($\delta t_\mathrm{acq}=150$~ms) allow to make another estimate with $v_d = 1000$~nm/s.

\paragraph{Case 3.}Fig.~\ref{fig:si:velocity}(e) shows an active island (labelled island~\#1) visible at the bottom of the image, and is observed at a different location in the following frame in Fig.~\ref{fig:si:velocity}(f). The displacement of island \#1 shown in Fig.~\ref{fig:si:velocity}(g) reveals a hopping length of $\ell = 360$~nm ($\delta t = 430$~ms and $\delta t_\mathrm{acq}=250$~ms). Note that the island does not show any smearing and its location is well defined, meaning that it was stationary during the acquisition in both images and as a consequence diffused during a maximum duration of $180$~ms across the two locations, allowing for an estimate $v_d = 1900$~nm/s.

\section{Thermal activation}
\label{si:thermalactivation}

\begin{figure}[t]
\centering
\includegraphics[width=\columnwidth]{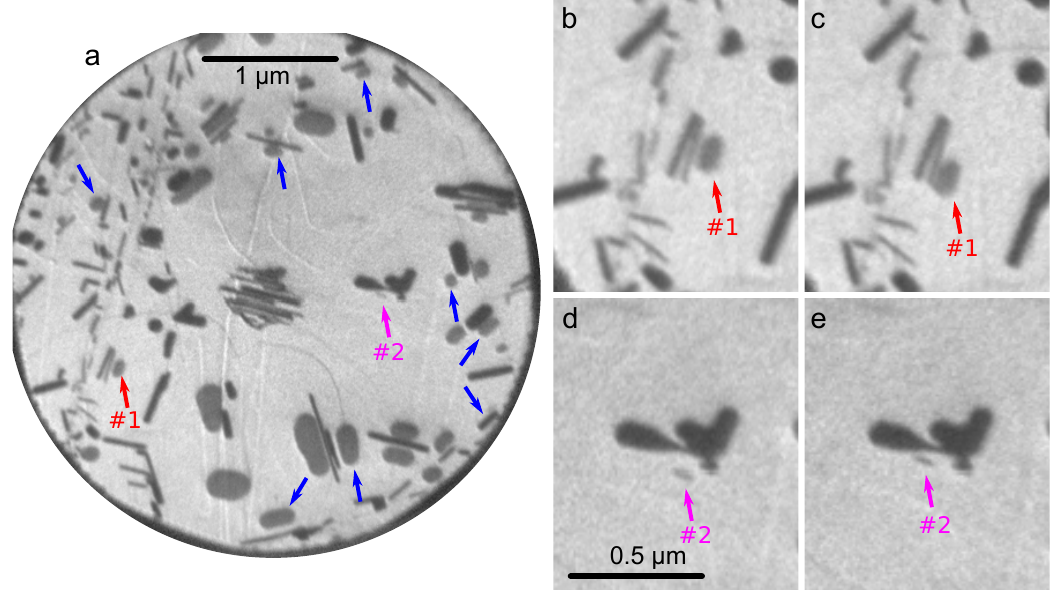}
\caption{\textbf{Thermal reactivation.} (a) Last frame of the LEEM sequence recorded at approximately 450~K (see MOV-3 for the corresponding LEEM movie). Arrows indicate several islands that become active after increasing the temperature. The red and pink arrows highlight islands \#1 and \#2 that are reactivated, shown in higher magnification in consecutive frames (b,c) and (d,e) respectively.}
\label{fig:si:thermal}
\end{figure}

Our observations show that in the initial stages of deposition, $\alpha$-Bi islands are more active and roughly $\sim20\%$ of them show spontaneous anomalous diffusion. However after some time, the islands typically reach a steady state and the hopping activity decreases, or vanishes completely. In order to verify whether it is possible to thermally reactivate these islands, we perform a LEEM acquisition with a mild increase of temperature, from room temperature ($\sim300$~K) to $400-450$~K. Figure~\ref{fig:si:thermal}(a) shows a LEEM image where several islands undergo an unpinning process by thermal activation.

\section{RI simulations}
\label{si:ri}

\begin{figure*}[t]
\centering
\includegraphics[width=\textwidth]{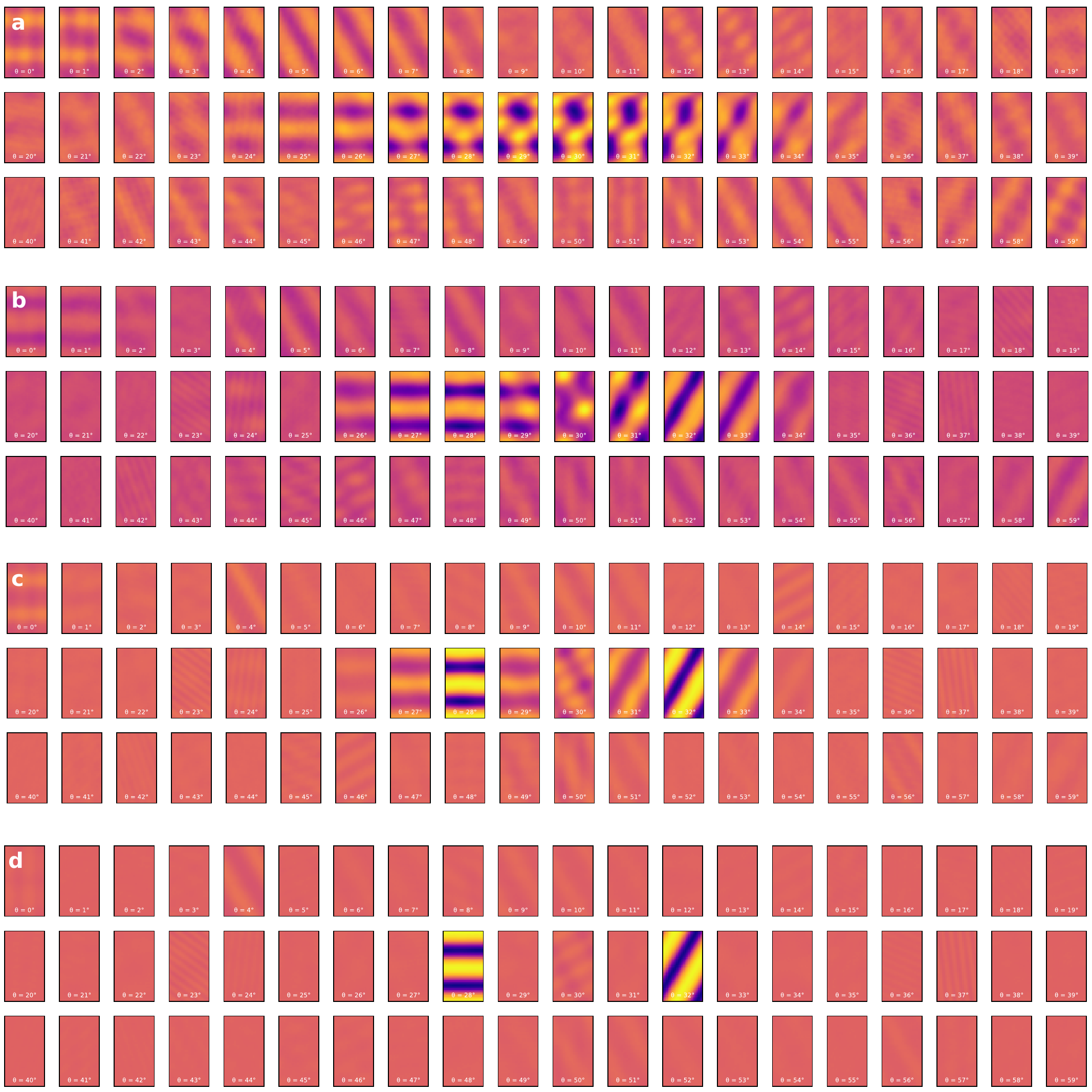}
\caption{\textbf{RI maps for different slab sizes.} RI maps of $\alpha$-Bi/HOPG for a varying twist angle $\theta=0^\circ$ to $\theta=59^\circ$ ($\Delta\theta=1^\circ$) obtained with different $\alpha$-Bi slab sizes: (a) $5\times5$ ($N=50$ atoms), (b) $10\times10$ ($N=200$ atoms), (c) $20\times20$ ($N=800$) and (d) $30\times30$ ($N=1200$ atoms).}
\label{fig:si_ri_theta}
\end{figure*}

\begin{figure}[t]
\centering
\includegraphics[width=\columnwidth]{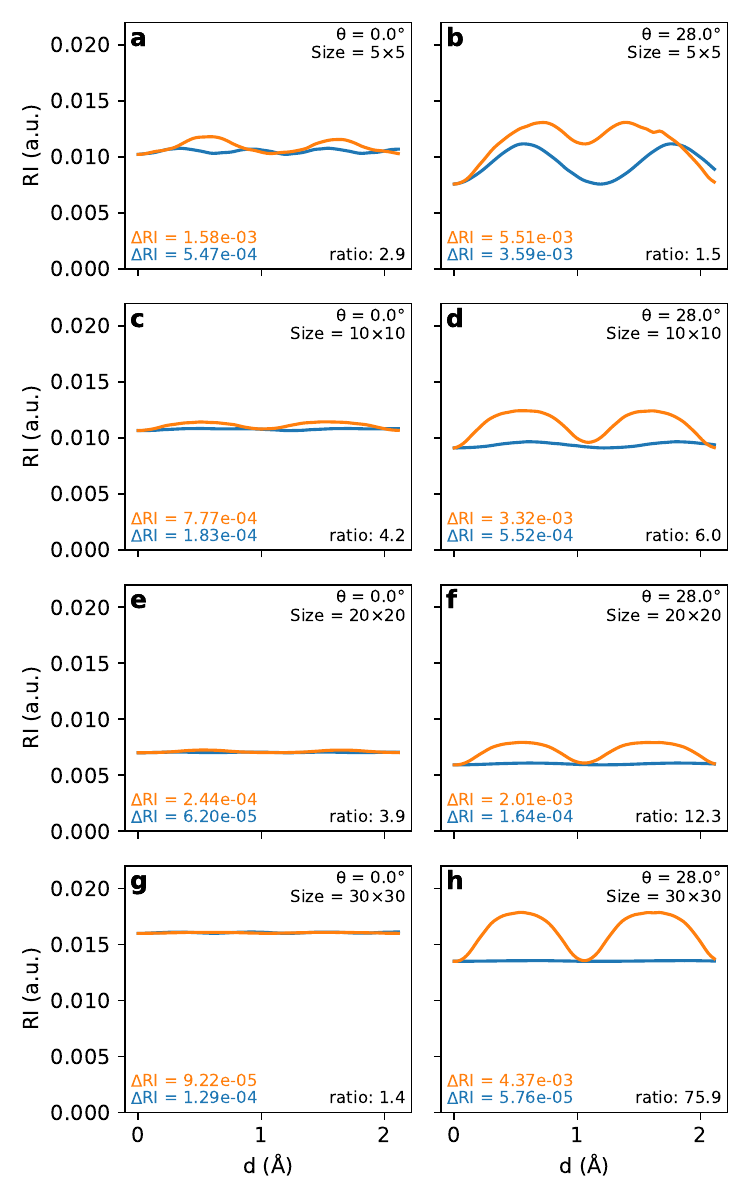}
\caption{\textbf{RI profiles for different slab sizes.} RI profiles across RI maps obtained for (a,b) $(5\times5)$, (c,d) $(10\times10)$, (e,f) $(20\times20)$ and (g,th) $(30\times30)$ $\alpha$-Bi slabs, for twist angles (a, c, e, g) $\theta=0^\circ$ and (b, d, f, h) $\theta=28^\circ$. The profiles are obtained along the graphite's zigzag (blue) and armchair (orange) directions (line profiles shown in Fig.~\ref{fig:si_ri_theta}).}
\label{fig:si_ri_corrugations}
\end{figure}

Registry index (RI) simulations \cite{Hod2012} (see Fig.~2(b) in the main text) allow to analyze the interaction between two crystalline bodies in contact in terms of relative twist and lateral translation. The RI is equal to the sum of the overlapping area between circles (representing atoms at the crystalline interface). The circles have a radius $r=0.3a$ where $a$ is the bond length ($a_{G}=1.42$~\AA, $a_{Bi}=3.05$~\AA).

The RI calculations were performed using different $\alpha$-Bi slab sizes ($5\times5$, $10\times10$, $20\times20$ and $30\times30$) where $r'_1\times r'_2 = (4.540\times4.828)$~\AA$^2$ on a graphite lattice ($r_1=r_2=2.461$~\AA). Refer to ball-and-stick models in Fig.~2(a) in the main text. Figure~\ref{fig:si_ri_theta} shows RI maps obtained for a large range of twist angles ($\theta=0$ to $\theta=59^\circ$ with $\Delta\theta = 1^\circ$), for three different $\alpha$-Bi slab sizes: $5\times5$, $10\times10$ and $30\times30$. For the smallest slab size in Fig.~\ref{fig:si_ri_theta}(a) the interlocking potential is minimized for $\theta\sim30^\circ$, however the RI corrugation for twist angles in this region are not unidirectional. The unidirectional character of the RI maps for these twist angles become visible for larger slabs ($10\times10$, see Fig.~\ref{fig:si_ri_theta}(b)), and is very pronounced for the $30\times30$ slab as shown in Fig.~\ref{fig:si_ri_theta}(d). This behaviour, where the potential energy landscape becomes more unidirectional with the adsorbate layer size, is in agreement with the theory \cite{Panizon2023}.

Fig.~\ref{fig:si_ri_corrugations} shows RI profiles obtained horizontally (graphite zigzag, blue) and vertically (graphite armchair, orange) across the RI maps shown in Fig.~\ref{fig:si_ri_theta}. The profiles are extracted from $\theta=0^\circ$ (left panels) and $\theta=28^\circ$ (right panels). The maximum corrugation along the profiles, $\Delta\mathrm{RI}$ are shown in all panels. Of particular importance are the blue plots for $\theta=28^\circ$ as they correspond to a measure of the translational energy landscape of $\alpha$-Bi along the \emph{nanohighway}. It is clear that the corrugation decreases with increasing the size of the slab. Furthermore the ratio $\Delta\mathrm{RI}_\mathrm{armchair}/\Delta\mathrm{RI}_{\mathrm{zigzag}}$ increases with the slab size, confirming the one-dimensional character of the low friction pathway. For twist angles characterized by full incommensurability such as $\theta=0^\circ$ in this case, the $\Delta\mathrm{RI}$ ratio (indicative of the friction anisotropy) tends to decrease with the slab size (and therefore becoming `more' isotropic), in agreement with fully incommensurate type-C contact \cite{Panizon2023}.

In general, the size-dependent simulations confirm the trend by which the one-dimensional character of the energy landscape is emphasized as the $\alpha$-Bi island size increases. These results agree with the theory \cite{Panizon2023}, although it is important to keep in mind that the model considers rigid lattices and ignores lattice relaxation, which becomes significant for large sizes in the case in a multitude of van der Waals heterostructures.


\section{Statistical tests}
\label{si:statistics}

\renewcommand\thefigure{S6}
\begin{figure*}[!ht]
\centering
\includegraphics[width=\textwidth]{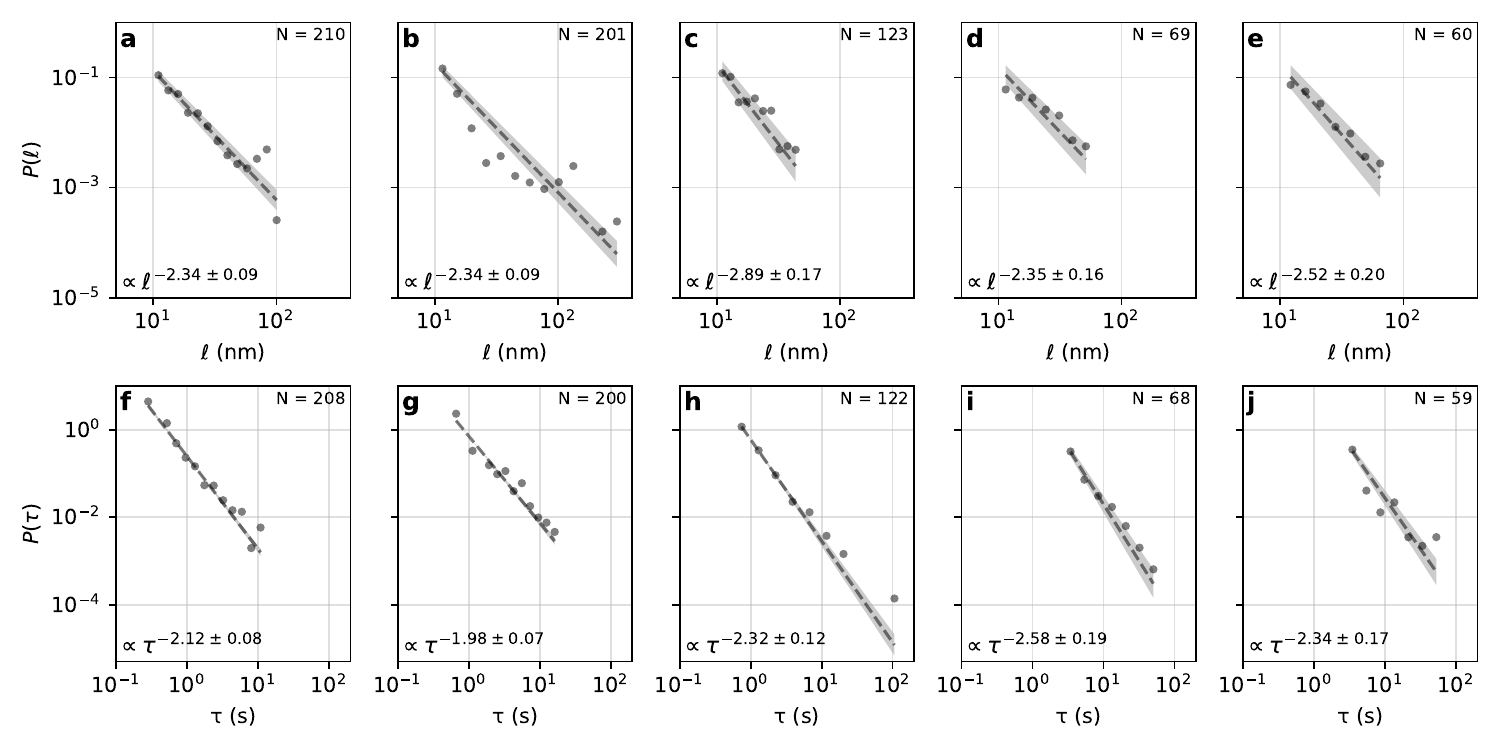}
\caption{\textbf{Statistics of hopping lengths and sticking times for individual islands.} Log-log histograms of (a-e) hopping lengths and (f-j) sticking times of individual islands with $N>58$ ($N$ number of events). Dashed lines correspond to power laws obtained via maximum likelihood estimation. The decay coefficients $\eta\pm\Delta \eta$ in of $P(\ell)$ and $P(\tau)$ are shown in the bottom left of each panel. The distributions are obtained from the same island in (a,f), in (b,g), in (c, h), etc.}
\label{fig:si_allislands}
\end{figure*}

Statistical analyses were performed on a total of 56 active $\alpha$-Bi islands from 9 movies containing over 8500 frames. Exponential fitting of area distributions (as shown in Fig.~3(c) in the main text) was performed via standard least-square methods; power-law fitting of hopping length and sticking time distributions ($\eta$ and $\Delta \eta$, see Fig.~3(a,b) in the main text) was performed using maximum likelihood estimation \cite{Clauset2009, Corral2019}. Due to pixel size, hopping lengths events $\ell<\Delta x=10$~nm and areas $A<100$~nm$^2$ are discarded from the analysis. The number of bins $n$ in histograms is $n = \lfloor\sqrt{N}\rfloor$ with $N$ the number of events.

Figure~\ref{fig:si_allislands} shows the distributions of both hopping lengths $\ell$ and sticking times $\tau$ for individual islands. Taken individually, the hopping lengths also follow the same trend as the overall population (data shown in Fig.~3 in the main text), \emph{i.e.}, $P(\ell)\sim\ell^{-\eta_\ell}$ and $P(\tau)\sim\tau^{-\eta_\tau}$ with \mbox{$1.98<\eta<2.89$}. The island-to-island variation is attributed to local inhomogeneities in the substrate, such as line or point defects, or due the presence of neighbouring islands in immediate proximity.


\section{Defect density}
\label{si:defects}

\renewcommand\thefigure{S7}
\begin{figure*}[t]
\centering
\includegraphics[width=\textwidth]{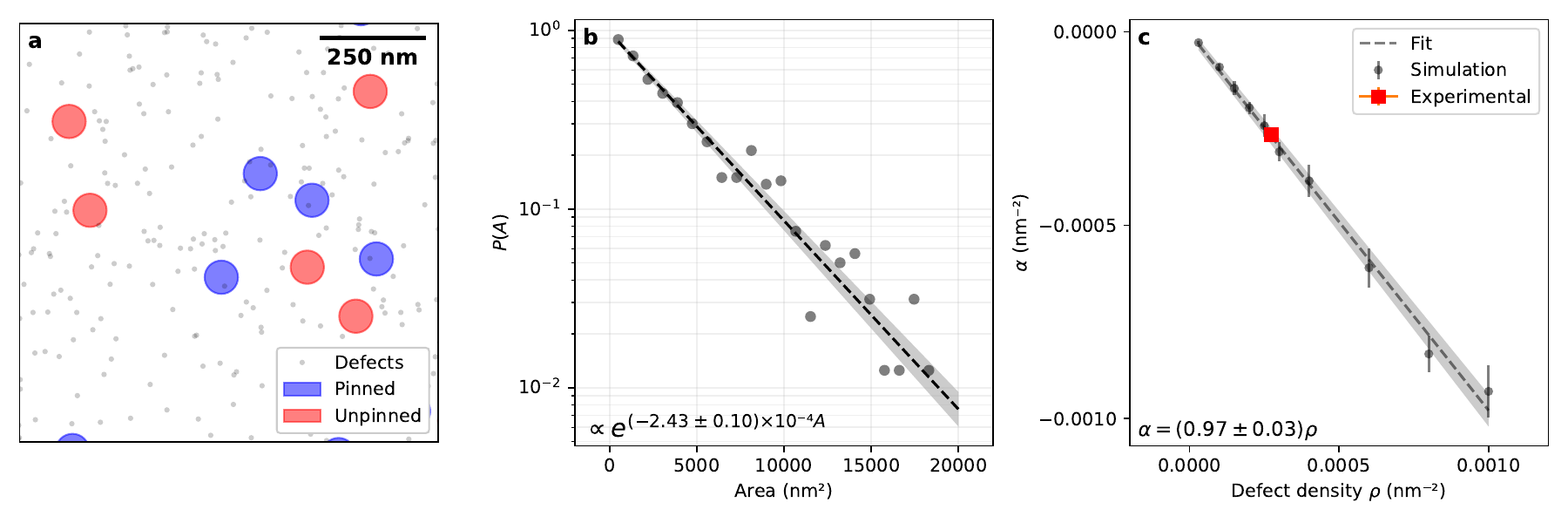}
\caption{\textbf{Substrate defect density and island pinning probability.} (a) Defect simulation example with defect density $\rho=2.5\times10^{10}$~cm$^{-2}$, island density of \mbox{$\varrho=1.0\times10^{9}$ cm$^{-2}$} and island area $A=5000$~nm$^{2}$ (blue: pinned, red: unpinned islands). (b) Probability of an island to be unpinned as a function of the area $P(A)$, in agreement with an exponential decay probability density function (dashed line shows the line of best fit $\exp(-\alpha A)$ (c) Exponential parameter $\alpha$ dependence as a function of defect density in the simulations, in very good agreement with $\alpha = 0.97\rho\simeq \rho$. The red square corresponds to the experimental observation (shown in Fig.~3(c) in the main text), predicting a defect density $\rho_{exp}=(2.74\pm0.16)\times10^{-4}$~nm$^{-2}$.}
\label{fig:si_defects}
\end{figure*}

In this section we investigate the role of defect density $\rho$ on the hopping probability $P$, as a function of the island area $A$. We develop and employ a simple model to gain insight on statistical relationship between island area and defect density, which we describe as follows. We consider a surface populated with a number of point defects located at $\mathbf{r}_j$, determined by the defect density $\rho$. On top of the surface, we randomly distribute a number of islands of area $A$ at location $\mathbf{R}_i$; for simplicity, we define the islands as circles. We fix the island density based on the experimental observations (excluding the islands decorating the terrace step edges, we estimate the island density $\varrho\simeq1\times10^{9}$~cm$^{-2}$). Once the defects and islands are generated, we evaluate the number of islands $N_{pin}$ that cover at least one point defect. This is achieved by evaluating the distance $d_{ij}=|\mathbf{R}_i-\mathbf{r}_j|$ between the island $i$ and defect $j$, 
allowing to determine if the island $i$ is pinned to defect $j$ by testing $d_{ij}<r$ with $r$ the radius of the circular island, $r=\sqrt{A/\pi}$. This model makes the assumption that the point defects are responsible for island pinning, and thus, if an island covers at least one point defect in the simulation, the island is considered immobilized. The probability of island pinning $\bar{P}$ in that context is given by:
\begin{equation}
\bar{P} = \frac{N_{free}}{N_{pin}}
\end{equation}
with $N_{free}$ the number of islands that do not overlap with a point defect ($N_{total}=N_{free}+N_{pin}$). The probability to find an active island (unpinned) is therefore given by:
\begin{equation}
P = 1-\bar{P} = 1-\frac{N_{free}}{N_{pin}}.
\end{equation}
The simulation consists of randomly positioning defects and islands based on their respective densities, and evaluating $P$ as a function of the area of the island $A$ (see Fig.~\ref{fig:si_defects}(a) for schematics). Intuitively, large islands are more likely to overlap with a point defect; conversely small islands are more likely to sit between defect sites promoting higher hopping activity; therefore $P(A)$ is a decreasing function. The simulation is run for a $4\times4$~$\mu$m$^2$ surface (\emph{i.e.}, for $N_{total}=160$ islands), with areas $A$ varying from $500$ up to $20000$ nm$^2$, as observed in our experiments.

Figure~\ref{fig:si_defects}(b) shows the probability $P(A)$ for an example defect density of $\rho = 2.5\times10^{10}$~cm$^{-2}$, which agrees very well with $P(A)\sim\exp(-\alpha A)$ with $\alpha=(2.43\pm0.10)\times10^{-4}$~nm$^{-2}$. Interestingly, the simulated probabilities are in very good agreement with the hopping probability $P(A)$ shown in Fig.~3(c) in the main text. For completeness, we run a series of similar simulations (with identical island density $\varrho$) this time with varying defect densities $\rho$ ranging from $3\times10^{-5}$ to $1\times10^{-3}$~nm$^{-2}$ (not shown). Figure~\ref{fig:si_defects}(c) shows the resulting decay coefficients $\alpha$ as a function of the defect density $\rho$. Interestingly, the defect density and the decay coefficient follow an identity rule (with a minus sign prefactor) with an excellent agreement, $\alpha/\rho = 0.97\pm0.03$. This allows to calculate the point defect density responsible for large island pinning, $\rho_{exp}=(2.32\pm0.11)\times10^{10}$~cm$^{-2}$.


\section{List of LEEM sequences}
\label{si:leemsequences}

The following list details the LEEM sequences mentioned in the manuscript (accessible in supplementary information and/or by request to the corresponding authors).

\begin{itemize}
\item{\textbf{MOV-1.} LEEM sequence ($\delta t = 2.75$ s) discussed in the main text and in Fig.~1 of the main text.}
\item{\textbf{MOV-2.} LEEM sequence ($\delta t = 0.577$ s) discussed in section~\ref{si:velocity} and pictured in Fig.~\ref{fig:si:velocity}.}
\item{\textbf{MOV-3.} LEEM sequence ($\delta t = 2.75$ s) recorded while increasing the temperature from room temperature to $\sim 400-450$~K. Discussed in section~\ref{si:thermalactivation}. Islands in the centre of the frame undergo ripening.}
\end{itemize}

\end{document}